\documentclass[final,5p,times,twocolumn]{elsarticle}
\usepackage{graphicx}
\usepackage{subcaption}
\usepackage{bm}
\usepackage{color}
\usepackage{amsmath}
\usepackage{float}
\usepackage{gensymb}
\usepackage{array}
\usepackage{tabularx}
\usepackage{multirow}
\usepackage{amssymb}
\usepackage{lipsum}

\begin{document}

\begin{frontmatter}



\title{Atomistic and experimental study of microstructural evolution in nanocrystalline iron subjected to irradiation}


\author[inst1]{Ivan Tolkachev\corref{cor1}}
\ead{ivan.tolkachev@eng.ox.ac.uk}
\author[inst2]{Daniel R. Mason}
\author[inst2]{Max Boleininger}
\author[inst2]{Pui-Wai Ma}
\author[inst1]{Daniel Long}
\author[inst3]{Eamonn T. Connolly}
\author[inst3]{Stephen P. Thompson}
\author[inst4]{Kenichiro Mizohata}
\author[inst1]{Felix Hofmann\corref{cor2}}
\ead{felix.hofmann@eng.ox.ac.uk}
\cortext[cor1]{corresponding author}
\affiliation[inst1]{organization={Department of Engineering Science, University of Oxford},
            addressline={Parks Road}, 
            city={Oxford},
            state={Oxfordshire},
            postcode={OX1 3PJ}, 
            country={United Kingdom}}
\affiliation[inst2]{organization={United Kingdom Atomic Energy Authority},
            addressline={Culham Science Centre}, 
            city={Abingdon},
            state={Oxfordshire},
            postcode={OX14 3DB}, 
            country={United Kingdom}}
\affiliation[inst3]{organization={Diamond Light Source},
            addressline={Harwell Science and Innovation Campus}, 
            city={Didcot},
            state={Oxfordshire},
            postcode={OX11 0DE}, 
            country={United Kingdom}}
\affiliation[inst4]{organization={Department of Physics, University of Helsinki},
            addressline={PL 43 (Pietari Kalmin katu 2)}, 
            city={Helsinki},
            postcode={00560}, 
            country={Finland}}

\begin{abstract}

Nanocrystalline materials have been proposed for use in future fusion reactors due to their high grain boundary density that may act as a sink for irradiation-induced defects. We use molecular dynamics to model collision cascades in nanocrystalline iron and compare the damage evolution to that observed in initially perfect, single crystalline iron. The nanocrystalline material is generated either by Voronoi tessellation or severe plastic shearing. Upon irradiation, the grains in nanocrystalline simulations coarsen, with all ultimately becoming single crystalline above 2 dpa. Above a damage dose of 1 dpa, nanocrystalline cells show a lower dislocation density and lower lattice swelling than their initially pristine counterparts. Experimental X-ray diffraction data is collected on nanocrystalline iron samples subjected to self-ion irradiation. Line profile analysis data agrees with the trends observed in the atomistic simulations, revealing the presence of an irradiation induced annealing process, with a clear reduction in micro-strain with increasing dose. We attempt to determine why some grains in our atomistic simulations grow, while others shrink, by creating a Toy Model that simulates volume exchange between grains based on different hypothesised exchange mechanisms. This suggests that irradiation-induced grain growth is consistent with random growth.

\end{abstract}




\end{frontmatter}




\section{Introduction}
\label{introduction}

During operation, deuterium-tritium fusion devices generate high-energy neutrons \cite{KURTZ201951} that penetrate the reactor material and interact elastically and inelastically with atomic nuclei, propagating up to tens of centimeters \cite{SATO2003501}. These interactions produce $\gamma$ photons \cite{RealiPhotons} and atomic recoils \cite{GILBERT2015121} which trigger collision cascades that form microstructural defects, for example dislocations and vacancy clusters. Such defects have been experimentally observed \cite{ElAtwaniSinks, HU2016235, DUBINKO2021105522} and extensively studied via molecular dynamics simulations \cite{MA2023154662, WU2023154653, BoleiningerCascade, GRANBERG2020151843, SAND201864}. Ultimately, these changes degrade the mechanical \cite{DUBINKO2021105522, SONG2024154998, SONG2024114144} and thermal transport \cite{REZA2022117926, HofmannThermal} properties of the material, shortening its lifespan.

Nanocrystalline (NC) materials, which are characterised by their small grain size ($<$ 100 nm) and hence large grain boundary (GB) density, are potentially attractive for fusion applications \cite{BEYERLEIN2015125}. Grain boundaries can act as sinks for material defects, whereby defects migrate towards, and are subsequently absorbed and annihilated by GBs \cite{ElAtwaniSinks, SamarasSinks, BaiSinks, TscoppSinks}, resulting in reduced defect density when compared with coarse-grained counterparts exposed to the same irradiation level \cite{NITA2004953, CHIMI2001355}. Furthermore, NC materials have been shown to possess enhanced mechanical properties \cite{WEI200677, MEYERS2006427, WangProperties}.

Song \textit{et al.} \cite{SONG2024114144} compared nanocrystalline (NC) and coarse-grained Eurofer97 under heavy iron ion irradiation. X-ray diffraction revealed irradiation-induced grain coarsening, aligning with earlier studies \cite{WANG198597, Atwater, Bufford, Kaoumi, NITA2004953}. In NC Eurofer97, dislocation density decreased, indicating an irradiation-induced annealing effect, while coarse-grained samples showed increased dislocation density upon irradiation. Using in-situ TEM, El-Atwani \textit{et al.} \cite{ELATWANI2019118} demonstrated that NC tungsten grain boundaries rapidly absorbed dislocation loops during irradiation. Similarly, Rose \textit{et al.} \cite{ROSE1997119} and Dey \textit{et al.} \cite{Dey} found lower defect densities in smaller-grained Pd and ZrO$_{2}$ samples post-irradiation.

Atomistic simulations have been used extensively to study the microstructural evolution of metals subjected to irradiation \cite{MA2023154662, BoleiningerCascade, GRANBERG2020151843, SAND201864, TscoppSinks, LeoMaPRM, DanielMason_diffusivity, DudarevCRA, LevoCascades, GranbergCascades}. Ma \textit{et al.} \cite{LeoMaPRM} utilised molecular dynamics (MD) simulations, implementing the creation relaxation algorithm (CRA) \cite{DudarevCRA}, to simulate radiation damage in NC tungsten. A noticeable grain growth was observed and the NC tungsten exhibited a significantly lower dislocation density than single crystal material. NC tungsten was also found to be more resistant to irradiation induced swelling than single crystalline cells, with volumetric swelling being $\sim$ 0.5 \% less at 2.5 dpa. Levo \textit{et al.} \cite{LevoCascades} performed collision cascade simulations on nanocrystalline nickel and similarly observed irradiation induced grain growth.

Severe plastic deformation (SPD) is an effective method for producing nanocrystalline (NC) materials, using techniques like high-pressure torsion, accumulative roll bonding, and equal channel angular pressing to reduce grain size with increasing strain \cite{AgarwalIOP, Borodachenkova17, ZHILYAEV2005277, ITO200932, STRANGWARDPRYCE2023101468}. The high grain boundary (GB) density in NC materials impedes dislocation motion, increasing yield strength via the Hall-Petch effect compared to coarse-grained materials. This is also accompanied by an increase in hardening, typically reducing ductility \cite{WangProperties, STRANGWARDPRYCE2023101468}. NC materials exhibit less irradiation-induced hardening \cite{ALSABBAGH2014} though the Hall-Petch effect only applies above 10 – 15 nm grain sizes; below this, the inverse Hall-Petch effect prevails \cite{SCHIOTZINVERSE}.

Molecular dynamics simulations have been used to study the creation of NC material through SPD \cite{GuoFeComp, TolkachevSims}. Guo \textit{et al.} \cite{GuoFeComp} used MD simulations to apply compressive strain to iron nanowires, resulting in the formation of NC iron. The creation of NC iron was associated to structural transition from body-centred cubic (BCC) to hexagonal close packed, followed by a reversal. Tolkachev \textit{et al.} \cite{TolkachevSims} induced grain refinement in pure iron through shearing of an initially pristine cell up to a shear strain of $\gamma = 10$. This study successfully produced nanocrystalline cells, and a number of distinct stages were observed with increasing shear strain: Highly distorted state and defect initiation; Nano-grain formation; Grain coarsening and rotation.

Reduced activation ferritic/martensitic (RAFM) steels are candidate structural materials for future fusion reactors and previous studies have shown them to possess an enhanced resistance to irradiation induced damage \cite{Garner_JNM_2000, Boutard, CABET2019510, MAZZONE2017655}. Garner \textit{et al.} \cite{Garner_JNM_2000} found that ferritic/martensitic steels, with body-centred cubic structure, offer better resistance to irradiation induced swelling than austenitic steels with a face-centred cubic (FCC) structure. In this study, we use pure iron as a model material for RAFM steels which are iron-based alloys. Studying iron is also useful due to its wide and varied use in other industrial settings.

Performing full collision cascade simulations involving many atoms still requires the use of large supercomputing resources \cite{LevoCascades, BoleiningerCascade, WU2023154653, GRANBERG2021153158} and full collision cascade simulations within nanocrystalline material are very limited. The work of Levo \textit{et al.} \cite{LevoCascades} utilised cells containing 446,000 atoms with 10 grains made by Voronoi tessellation \cite{Voronoi1908}. This means that the nanocrystalline cells contained grain boundaries but no other defects were seeded prior to irradiation. Damage levels in the range 0.2 dpa to 1 dpa \cite{LevoCascades, MA2023154662} were considered in this study. This leaves us with two major questions: Is the presence of crystal defects in nanocrystalline material important for the subsequent microstructural evolution? How does the material structure evolve beyond 1 dpa? 

We aim to address these questions and seek to gain insight into the mechanisms that control microstructural evolution in nanocrystalline materials exposed to irradiation. We initially create nanocrystalline cells through Voronoi tessellation and through the high strain method developed in \cite{TolkachevSims}. We also consider pristine iron cells to facilitate comparison between nanocrystalline and single crystal material. Our simulation cells contain 1,024,000 atoms. Collision cascade simulations are used to mimic radiation damage under athermal conditions within our material, and follow the method laid out in \cite{BoleiningerCascade} from 0 to 5 dpa. The results of the collision cascade simulations are compared to a line profile analysis of experimentally obtained X-ray diffraction profiles of high-pressure-torsioned and subsequently self- ion irradiated iron discs.

\section{Methods}

\subsection{Simulation Setup}

All molecular dynamics simulations were performed using LAMMPS \cite{LAMMPS}. NC simulation cells were generated via Voronoi tessellation \cite{Voronoi1908}, which creates equiaxed grains. Two grain sizes were used: cells seeded with 5 grains (d$_{c}$ = 16.6 nm) or 20 grains (d$_{c}$ = 5.2 nm).  For each size, five independent Voronoi cells were created, with results averaged and uncertainties calculated across these samples. Additionally, five more cells were generated using the SPD method from \cite{TolkachevSims}, which generates NC material via high shear strain deformation. Finally, five single crystal pristine iron cells were created for comparison with the NC material.

Ferenc \textit{et al.} \cite{FERENC2007518} approximate the probability distribution for the volume, $V$, of polyhedra created through Voronoi tessellation as

\begin{equation} \label{eq1}
    p(y) = \frac{3125}{24} y^4 \exp(-5y)
\end{equation}
\\

\begin{figure}
\begin{subfigure}{0.48\textwidth}
\includegraphics[width=\linewidth]{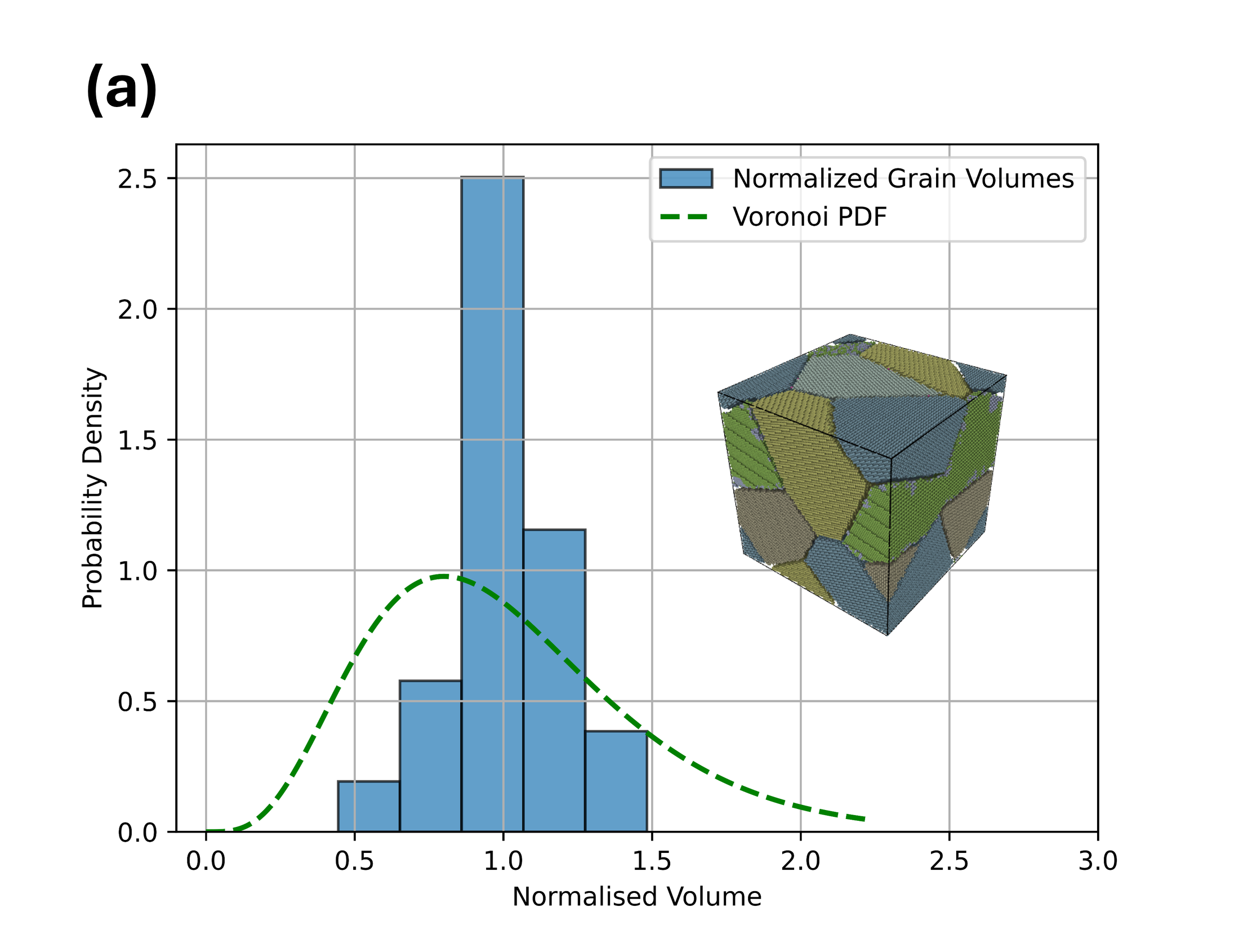}
\label{fig:Fig1a}
\end{subfigure}

\medskip
\begin{subfigure}{0.48\textwidth}
\includegraphics[width=\linewidth]{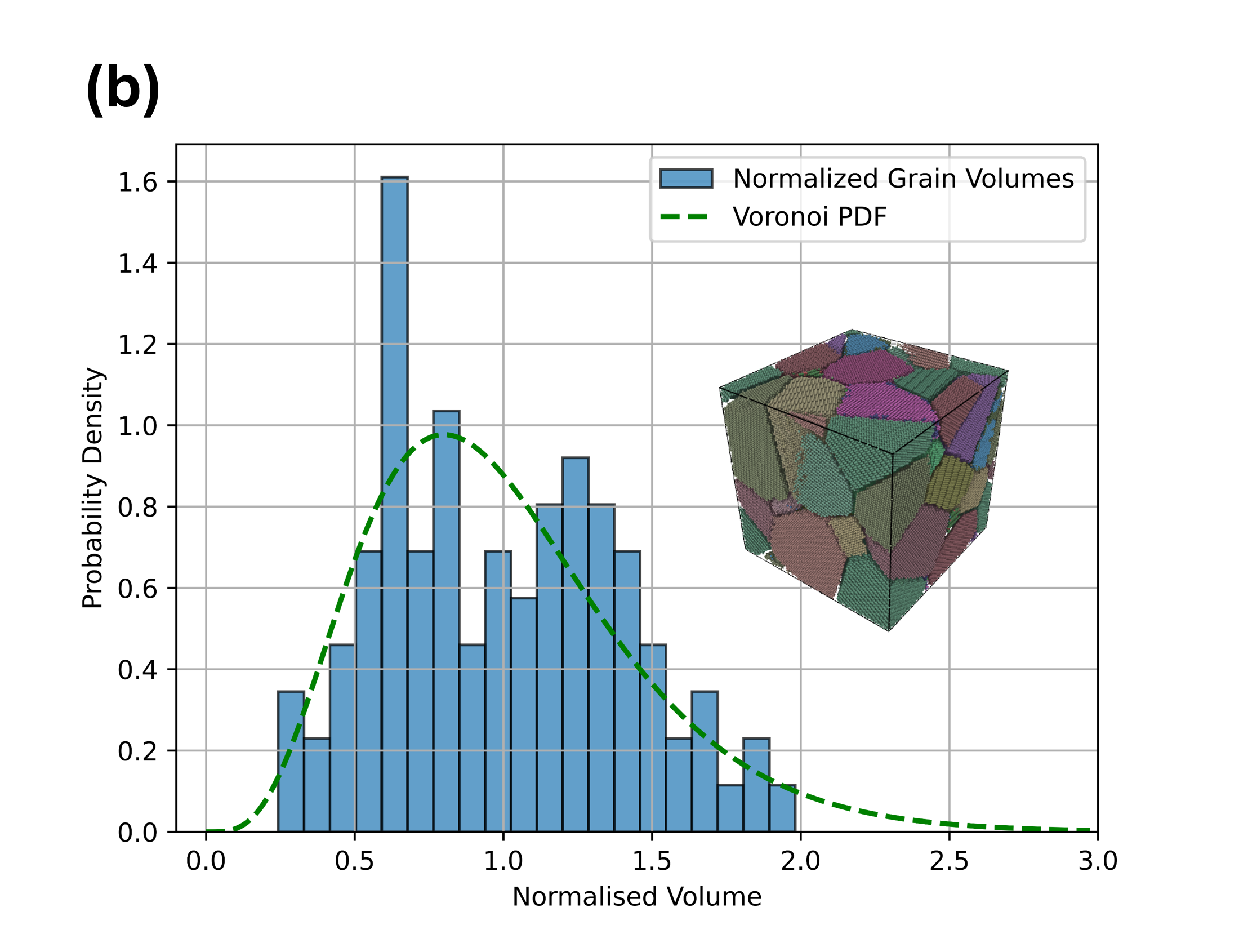}
\label{fig:Fig1b}
\end{subfigure}

\medskip
\begin{subfigure}{0.48\textwidth}
\includegraphics[width=\linewidth]{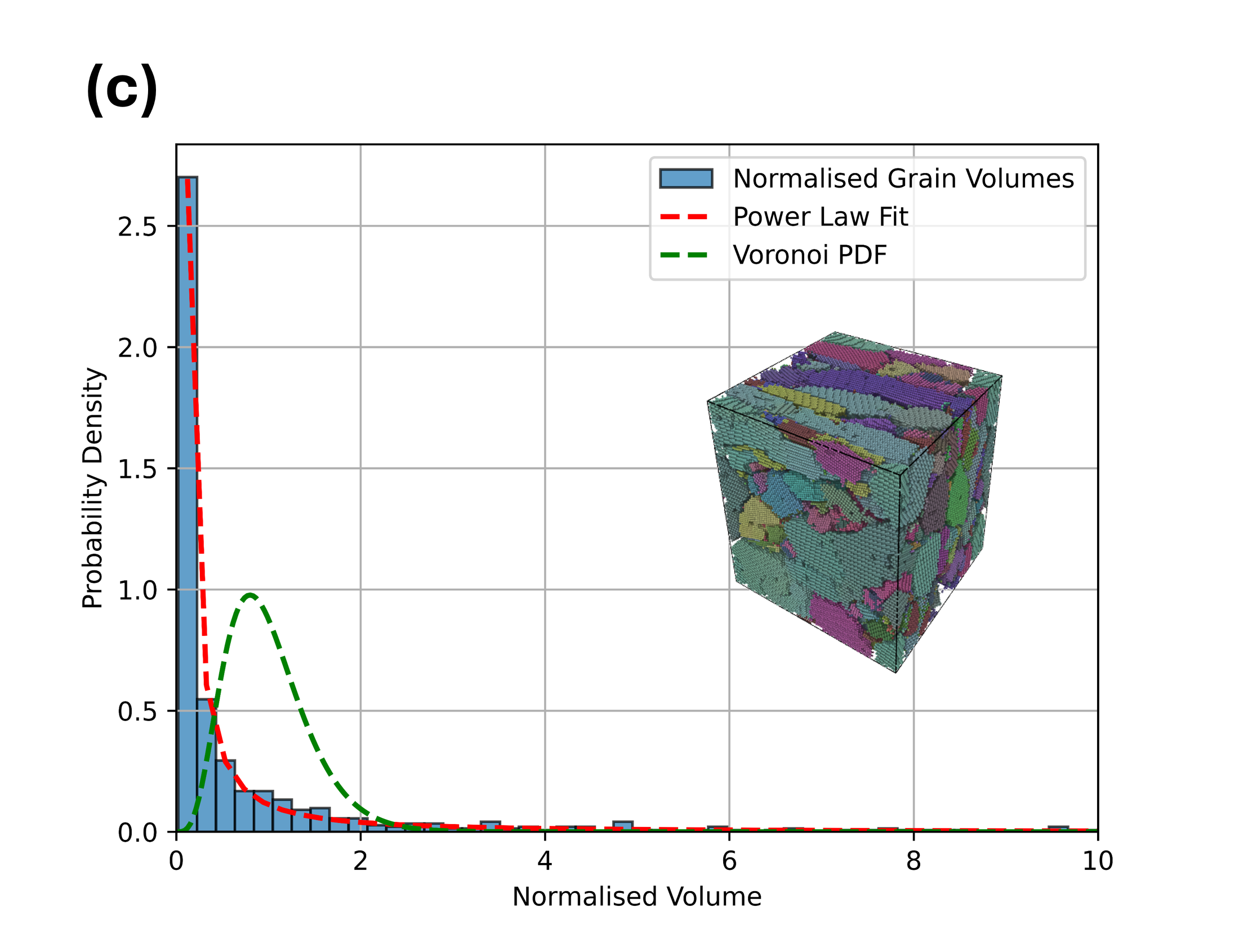}
\label{fig:Fig1c}
\end{subfigure}

\caption{Normalised volume against probability density for the Voronoi and Sheared cells. The green dashed line relates to the PDF in Eq. (\ref{eq1}) and the red dashed line relates to the PDF in Eq. (\ref{eq2}); (a) d$_{c}$ = 16.6 nm, (b) d$_{c}$ = 5.2 nm, (c) \textit{sheared} cells.} \label{fig:Fig1}
\end{figure}

In this equation, $y = V / \langle V \rangle$, is the \textit{normalised volume}. In Figure \ref{fig:Fig1}, the normalised volume is plotted against the probability density for the nanocrystalline cells. The starting volumes are summed for each individual condition, and the Voronoi probability distribution function (PDF) is plotted as a green dashed line. The number of bins is selected to be the average number of grains in each simulation for each condition. It is difficult to ascertain whether the d$_{c}$ = 16.6 nm case is well correlated to Eq. (\ref{eq1}) due to the low availability of data, however, Fig. \ref{fig:Fig1}(b) shows that the d$_{c}$ = 5.2 nm case is well correlated with the Eq. (\ref{eq1}) approximation.

It is noticeable however, that the size distribution of grains for the sheared cells is not well correlated to the Ferenc \cite{FERENC2007518} approximation, as shown by the green dashed line in Figure \ref{fig:Fig1}(c). Instead, a power law was most suitable to describe the volume distribution in the case of the sheared cells, shown with a red dashed line. This was found to take the form of a Zipf-Mandlebrot distribution \cite{ZipfMandlebrot} given by

\begin{equation} \label{eq2}
    p(y) = \frac{(c-1)}{b}\left(\frac{y}{b} + 1\right)^{-c}
\end{equation}
\\

where b = 0.1 and c = 1.51. All of the cells contain 1,024,000 atoms which, in a perfect crystal, correlates to 80 x 80 x 80 unit cells, with each cell containing two atoms. This corresponds to a 22.84 x 22.84 x 22.84 nm cell. The lattice parameter for iron was chosen as a = 2.855 {\AA} in accordance with the Mendelev \textit{et al.} \cite{MENDELEVPOT} Fe2 potential, which is used in this study. This is a well established embedded-atom method (EAM) potential belonging to the Ackland-Mendelev family of iron potentials that have been used extensively to study irradiation damage in iron and iron alloys \cite{SAND201864, DudarevCRA, KEDHARNATH2019444, YE2021152909, MALERBA201019}. The work of Malerba \textit{et al.} \cite{MALERBA201019} confirmed the transferability of the Ackland-Mendelev iron potentials for use in collision cascade simulations when directly compared to density functional theory.

The Voronoi and pristine cells were all relaxed using the Fe2 \cite{MENDELEVPOT} potential, with the cell preparation method following that of \cite{LeoMaPRM}. The SPD cells, denoted as the \textit{sheared} cells, were initially thermalised to 300 K and then sheared up to $\gamma = 10$. After this point, the same procedure as for the Voronoi and pristine cells was followed, where the cells were allowed to achieve stress free conditions at a temperature of 0 K. 

\subsection{Collision Cascade Simulations}

To simulate irradiation damage in the NC and pristine cells, the collision cascade method proposed by Boleininger \textit{et al.} \cite{BoleiningerCascade} was followed. The current study only considers the athermal evolution of the iron microstructure subjected to collision cascades. Due to the limitations of MD simulations, the dose rate in these cascade simulations is very high ($d\phi/dt \sim 1 \ dpa/\mu s$), which means that there is insufficient time for thermally activated processes that contribute to damage recovery \cite{BoleiningerCascade}. It is in this athermal regime that successive cascade simulations offer a credible depiction of a heavily irradiated material \cite{BoleiningerCascade}. Studying this regime is appropriate for regions of a fusion device which are actively cooled, e.g. near a cooling pipe, and for direct comparison to room temperature experiments.

The generation of radiation defects was modelled by explicitly tracing atomic trajectories from the moment a high-energy particle strikes an atom within a crystal lattice, commonly referred to as the recoil or primary knock-on atom (PKA). If sufficient kinetic energy, $E_{R}$, is transmitted from the high-energy particle to the recoil atom, it is displaced from its lattice site initiating a cascade of atomic collisions within the material. 

To simulate PKA production from high-energy neutrons, a small number of atoms $N_{c}$ were randomly assigned recoil energies, obtained from SRIM \cite{ZIEGLER20041027}. SRIM was used to model the implantation of 20 MeV Fe ions into pure iron for 10,000 ions, producing a recoil energy probability distribution (50 eV–20 keV) used in the MD simulations (see Appendix A). A displacement energy of 40 eV was used \cite{GRANBERG2020151843, STOLLER201375}, so the minimum damage energy was taken as 50 eV. To avoid cascade fragmentation, which occurs above 30 keV in iron \cite{De_Backer_2016}, the maximum recoil energy was limited to 20 keV.

To ensure that there is no cascade overlap, the exclusion radius method was used. Cascade positions are selected at random and iteratively checked for overlap, whereby the positions will only be accepted if all the distances between cascade positions exceed the exclusion distance. The number of cascades modelled per step were selected based upon the increment of damage required in each step. In this study, an increment of $\sim$ 0.0002 displacements per atom (dpa) was selected between consecutive cascade runs. The PKA energies were converted to damage energies, $T_{d}$, through the Lindhard \cite{Lindhard} model and the Norgett-Robinson-Torrens (NRT-dpa) model \cite{NRT1,NRT2} was subsequently used to determine the number of atomic displacements, $N_{d}$, generated by any given cascade.




During each cascade run, atoms evolved in an NVE ensemble for at least 5 ps until the cell cooled below 100 K. Following \cite{BoleiningerCascade}, a frictional force was applied to atoms with kinetic energy above 10 eV to model energy loss from electronic excitation. An additional damping term accounted for electron-phonon coupling \cite{Mason_2015} in atoms below their melting temperature. After each run, atomic velocities were set to zero and positions relaxed via conjugate gradient minimization. The simulation cell and box stresses were also relaxed. Cascade runs continued in $\sim$ 0.0002 dpa increments until reaching a target dose of 5 dpa for all cells.

These simulations were found to be very repeatable. We refer to Appendix B for the potential energy within each simulation cell type for each of the five independent cases. The potential energy evolution follows nearly identical trajectories for a given simulation type.

\subsection{Experimental Sample Preparation}

The high purity iron used in this work was manufactured under the European Fusion Development Agreement (EFDA) programme (contract EFDA-06-1901), and the chemical composition of the as-delivered iron can be found in \cite{EFDA1, EFDA2}. The starting material had a chromium content of $< 0.0002 \ wt\%$, carbon content of $4 \ wppm$, sulfur content of $2 \ wppm$, oxygen content of $4 \ wppm$, and nitrogen content of $1 \ wppm$. The mean as-delivered grain size was also previously found \cite{SONG2024154998}, through electron backscatter diffraction (EBSD), to be $187 \pm 150 \ \mu m$.

\begin{figure}
\includegraphics[width=\linewidth]{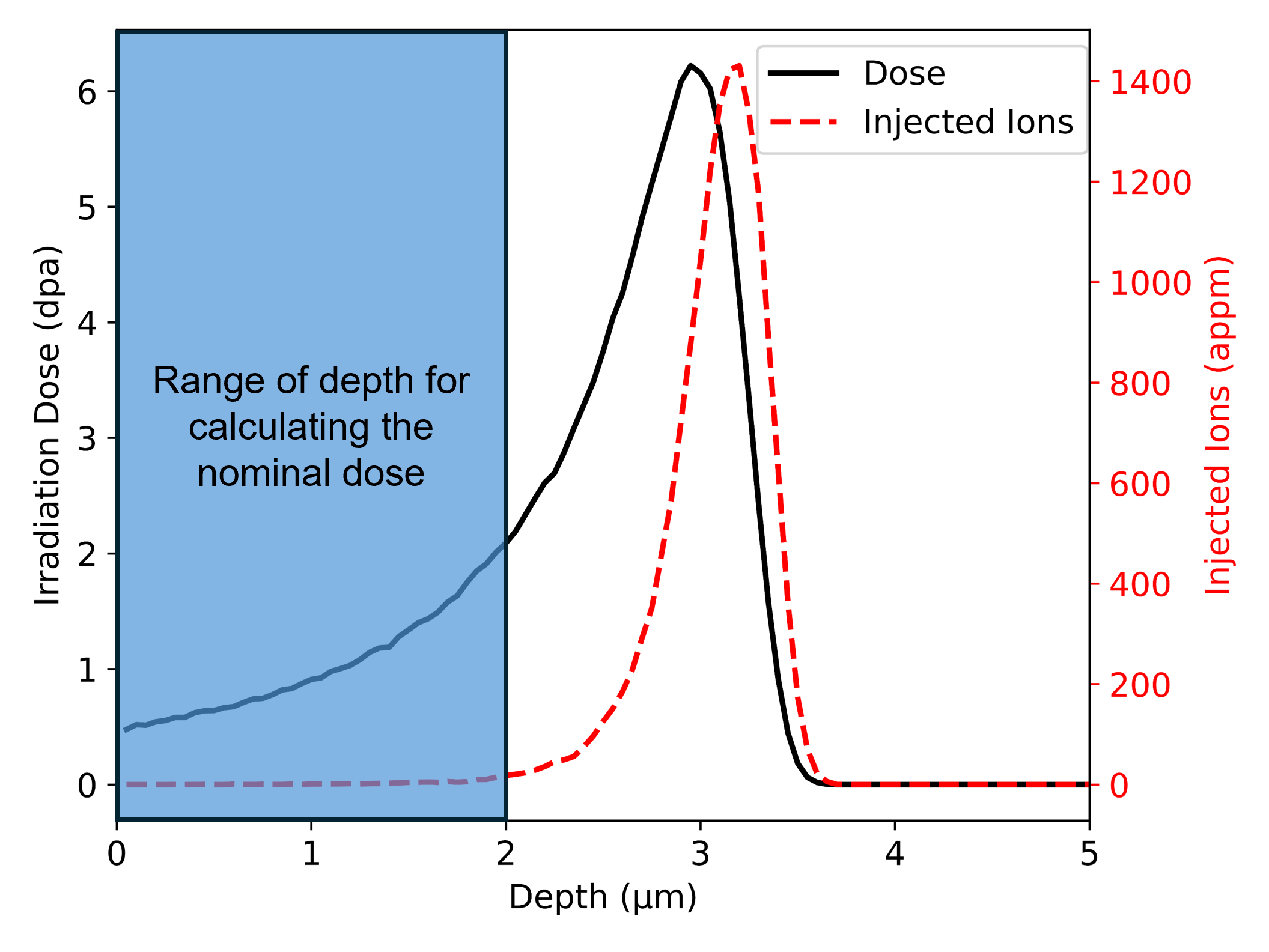}
\caption{Damage profile as a function of depth calculated for 1 dpa using SRIM \cite{ZIEGLER20041027} \textit{vacancy.txt} method \cite{STOLLER201375, AGARWAL202111} for 20 MeV $Fe^{4+}$ ions into Fe.} \label{fig:Fig2}
\end{figure}

Iron samples were cut into 5 mm diameter, 1 mm thick discs and processed via high-pressure torsion (HPT) using a Zwick Roell Z100 Materials Testing Machine in quasi-constrained mode. Each sample was compressed with 80 kN ($\approx$ 4 GPa) and subjected to 30 torsional turns at $5 \degree s^{-1}$. This ensured a saturated microstructure beyond $\sim$ 0.5 mm from the center, as saturation in iron occurs at $\sim$ 100 strain \cite{TEJEDOR2019597, 2009MD200812}. All samples were then mechanically ground with SiC paper and polished with diamond suspension.

The HPT-processed iron discs were irradiated at the Helsinki Accelerator Laboratory using 20 MeV Fe$^{4+}$ ions at a constant temperature of 300 K under vacuum. The dose profile was calculated with the Quick K-P model in SRIM \cite{ZIEGLER20101818} using the \textit{vacancy.txt} method detailed in \cite{STOLLER201375, AGARWAL202111}, with a displacement energy of 40 eV for iron \cite{GRANBERG2020151843, STOLLER201375}. Figure \ref{fig:Fig2} shows the damage profile and ion distribution for the 1 dpa case. The nominal dose refers to the average dose within the first 2 $\mu$m of the sample, where ion concentration is low and damage reasonably uniform. Dose is assumed to scale linearly with ion fluence. Nanocrystalline iron samples were irradiated at nine nominal doses: 0.001, 0.003, 0.01, 0.03, 0.1, 0.3, 1, 2.5, and 5 dpa, with an unirradiated sample retained for comparison.

\subsection{Grazing Incidence X-ray Diffraction}

Grazing incidence X-ray diffraction (GIXRD) measurements were used to investigate changes in grain size and dislocation density within the irradiated iron discs. This technique allows us to control the thickness of the layer of material that is probed, ensuring that the line profiles are dominated by the top 2 - 3 $\mu$m layer of each sample so that only the irradiated layer is probed and the contribution from the unirradiated bulk is removed. The measurements were carried out at Beamline I11 at the Diamond Light Source, which utilises a high-resolution multi-analyser crystal (MAC) detector. An X-ray energy of 15 keV ($\lambda$ = 0.824570 $\AA$) was utilised, with a grazing incidence angle of 7 $\degree$ (attenuation depth $\sim$ 2.74712 $\mu$m).

To determine the instrumental contribution to the powder diffraction patterns and calibrate experimental geometry, an NIST Silicon 640C sample was measured as a calibration standard. This was used to fit the instrument broadening function proposed by Caglioti \textit{et al.} \cite{CAGLIOTI1958223}. Full details of the instrumental broadening contribution can be found in Appendix C.

\subsection{Computational Analysis Methods}

OVITO \cite{Stukowski_2010} was used for analysis in this work. The polyhedral template matching (PTM) algorithm \cite{Larsen_2016} identified local atomic orientations, which were used in the grain segmentation modifier for grain identification. The PTM modifier identifies local crystallographic orientations in terms of a quaternion, and each quaternion can be projected into Rodrigues space \cite{DAI2015144}. An RGB colouring scheme can be mapped onto each orientation allowing each atom to be coloured acording to the local crystallographic orientation \cite{ALBOU20103022}. The dislocation analysis modifier (DXA) \cite{Stukowski_2012} is used to detect dislocation lines.

The method of grain identification proposed by Ma \textit{et al.} \cite{MA2023154662} was also used for comparison. This method identifies the local environment for the atoms within the cell and assigns those atoms into grains based on comparable orientations. Identification of grain boundaries was also possible using the method proposed by Ma, which allows the computation of the distance to the nearest grain boundary for the atoms within each grain. The dislocation density inside grains was estimated by removing the atoms close to grain boundaries ($d<4$ {\AA}) and then using DXA for dislocation identification. 

The method proposed by Ma \textit{et al.} \cite{MA2023154662} also calculates the local crystallographic orientation of atoms closest to the x = [100], y = [010], and z = [001] set of axes. These atomic orientations were subsequently projected onto a geodesic sphere with equal areas and the density of the projected orientations in each area is calculated. This allowed for the creation of pole figures which were weighted by the number of atoms with certain crystallographic orientations, allowing the identification of dominant orientations within the cells.

To facilitate direct comparison between the XRD data obtained in experiments and the MD simulation, line profiles are calculated for the MD cells using the Debye scattering equation \cite{DebyeScattering}. These line profiles are analysed utilising the method proposed by Williamson and Hall \cite{WILLIAMSON195322} (denoted as \textit{Williamson-Hall}), described below, which enables direct comparison of experimental and computational data.

\subsection{Experimental Analysis Methods}

To analyse the recorded powder diffraction profiles, the Williamson-Hall \cite{WILLIAMSON195322} method was utilised. This is a well established method for determining the average grain size, $D$, and microstrain, $\epsilon$, in the samples through analysis of X-ray line profiles.

\section{Results}

\subsection{Initial Nanocrystalline Structure}

\begin{figure*}[t]
\includegraphics[width=\linewidth]{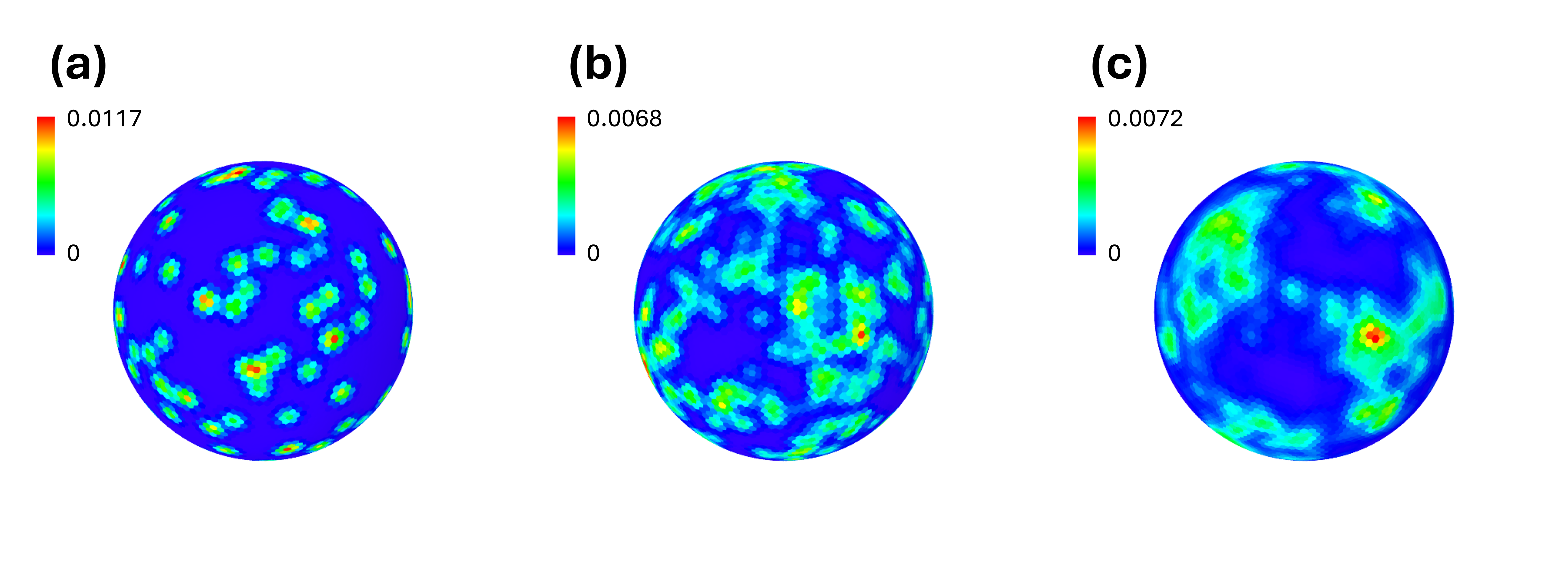}
\caption{[001] pole figures for the simulation cells prior to irradiation. Figures are coloured by density with the five distinct cells being summed: (a) d$_{c}$ = 16.6 nm; (b) d$_{c}$ = 5.2 nm; (c) \textit{sheared} cells.}
\label{fig:Fig3}
\end{figure*}

Figure \ref{fig:Fig3} shows the [001] pole figures for the NC simulation cells prior to irradiation. Each pole figure is the sum of orientations for each starting condition, with Figure \ref{fig:Fig3}(a) showing the d$_{c}$ = 16.6 nm cells, Figure \ref{fig:Fig3}(b) showing the d$_{c}$ = 5.2 nm cells, and Figure \ref{fig:Fig3}(c) showing the \textit{sheared} cells. 

Cells that were created through Voronoi tessellation appear to have no distinct texture, with many different patches of orientations being present in both Figures \ref{fig:Fig3}(a) and \ref{fig:Fig3}(b). However, there does appear to be a preferred orientation in the \textit{sheared} cells in Figure \ref{fig:Fig3}(c), where there is some evidence of orientation clustering. 

\begin{figure}
\begin{subfigure}{0.43\textwidth}
\includegraphics[width=\linewidth]{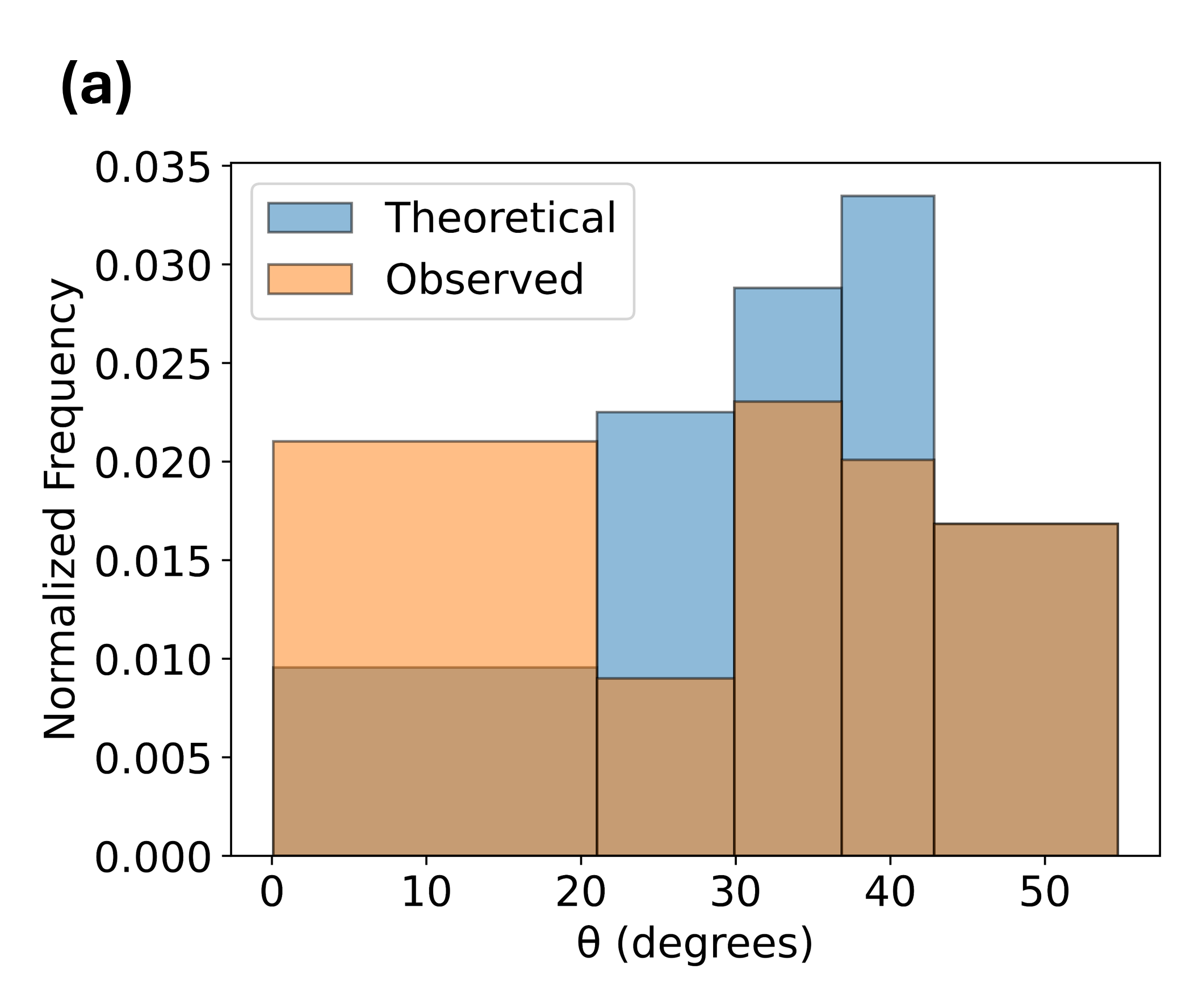}
\label{fig:Fig4a}
\end{subfigure}

\smallskip
\begin{subfigure}{0.43\textwidth}
\includegraphics[width=\linewidth]{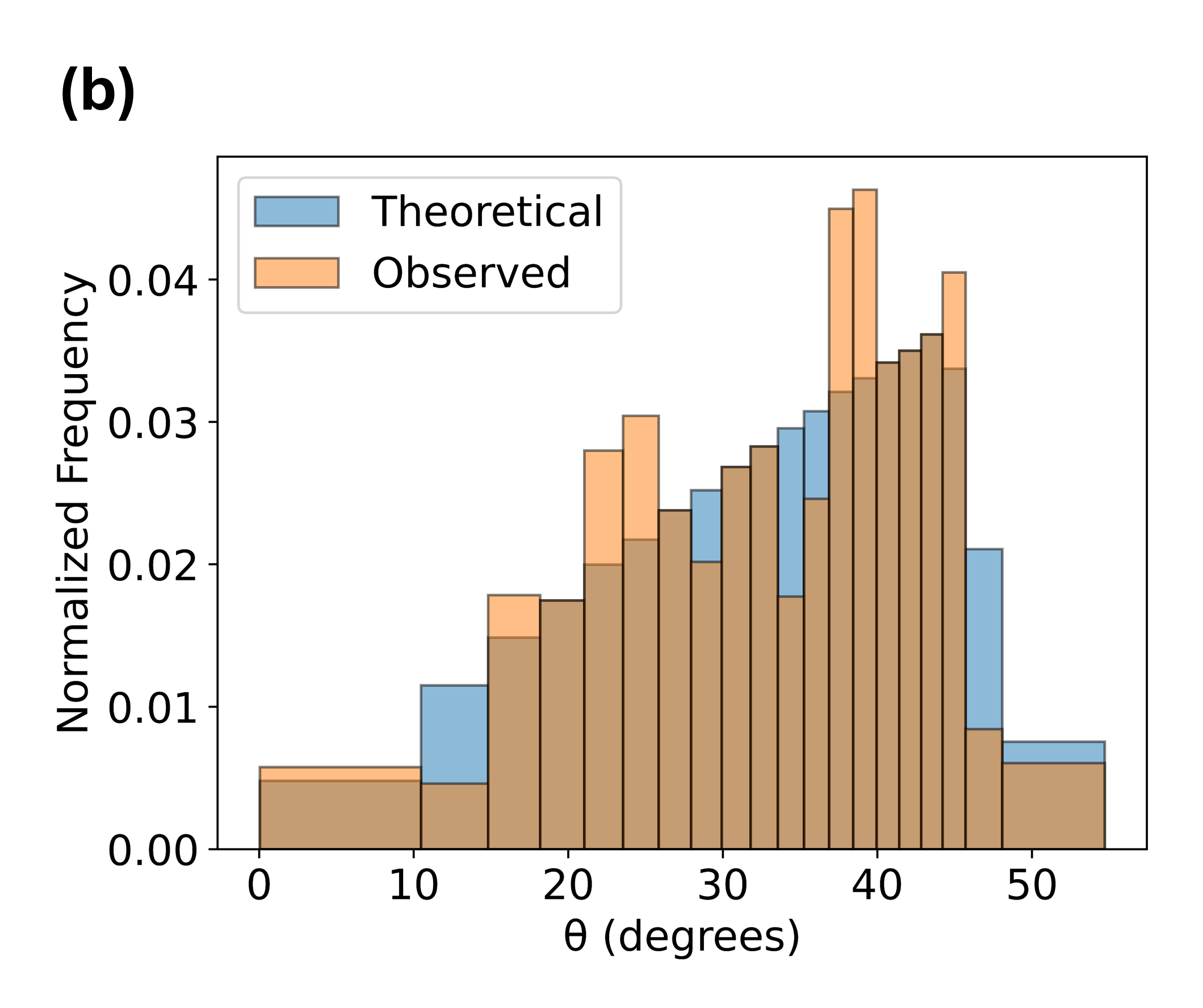}
\label{fig:Fig4b}
\end{subfigure}

\smallskip
\begin{subfigure}{0.43\textwidth}
\includegraphics[width=\linewidth]{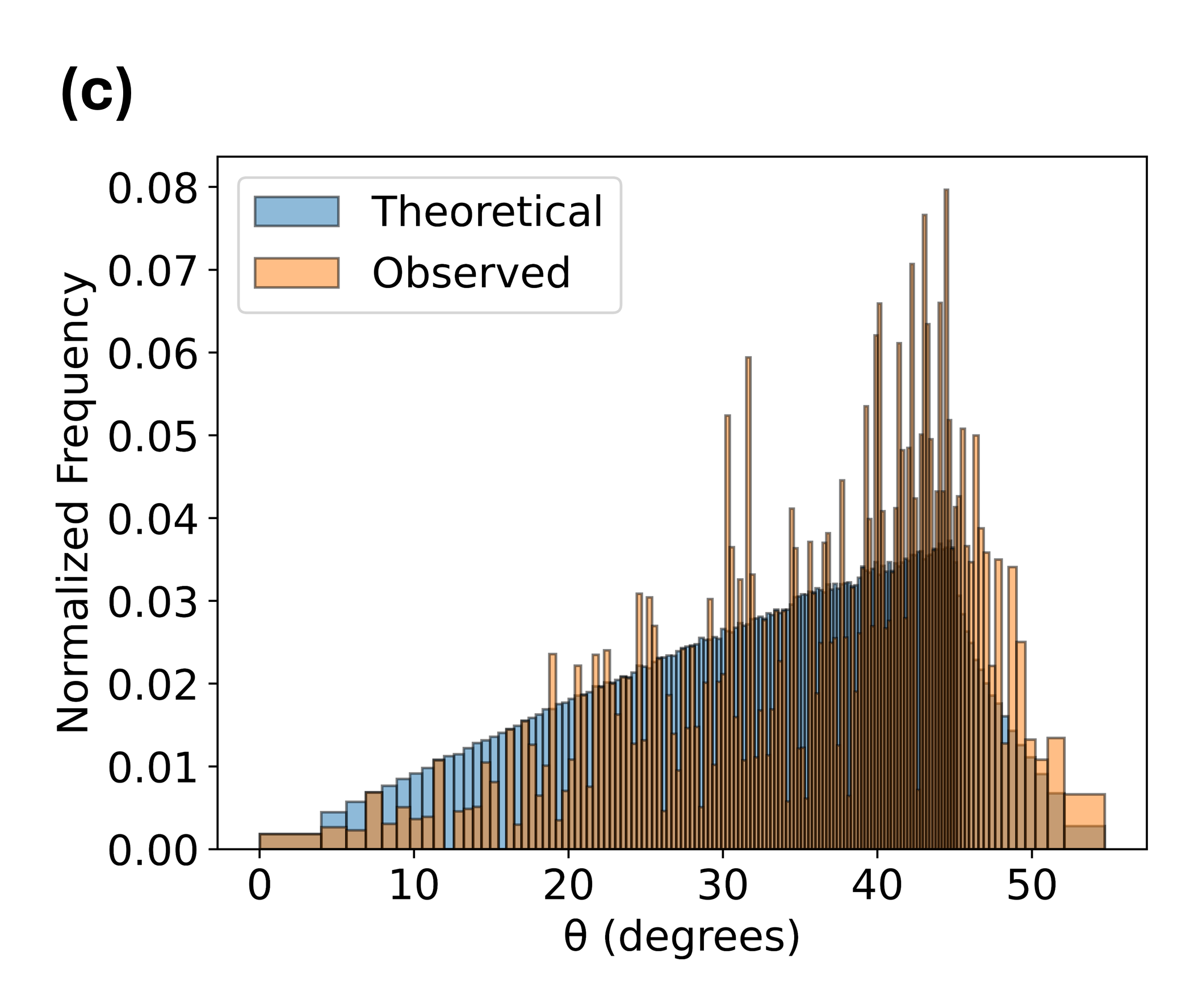}
\label{fig:Fig4c}
\end{subfigure}

\caption{Histograms of normalised frequency of grain orientations against orientation angle for the nanocrystalline simulations. The theoretical (blue) bins represent a truly random distribution and the observed (orange) bins represent the grain orientations found at 0 dpa. The number of bins is the average number of grains in each distinct simulation; (a) d$_{c}$ = 16.6 nm, (b) d$_{c}$ = 5.2 nm, (c) \textit{sheared} cells.}\label{fig:Fig4}
\end{figure}

A statistical analysis is carried out to determine whether the initial grain orientations in the Voronoi and sheared cells are indeed randomly distributed. Figure \ref{fig:Fig4} shows the theoretical and observed distributions of grain orientations in the nanocrystalline cells. The theoretical distribution is produced by randomly generating 1 million quaternions, converting them to rotation matrices, and applying those matrices to a x = [100], y = [010], and z = [001] coordinate system. The vectors closest to the original z-axis are found, and $\theta$ is the angle between these vectors and z = [001]. Note, that the maximum angle of $\theta$ is 54.7 degrees, as any vector which has a $\theta$ beyond this value would no longer be the closest to the original z-axis due to cubic symmetry.

The distribution of grain orientations in the MD cells are also computed in the same way. Figure \ref{fig:Fig4} shows the observed orientations as a sum of all five distinct simulations for each starting condition. Dynamic binning is employed to ensure that there are at least 5 observed data points in each bin. The bin widths were calculated using the theoretical data.

$\chi^{2}$ tests are performed by comparing the theoretical and observed orientations in Figure \ref{fig:Fig4}. The significance level, $\alpha$, is taken to be 0.05. Figure \ref{fig:Fig4}(a) shows the grain orientations for d$_{c}$ = 16.6 nm case (5 Grain Voronoi cells). The number of bins is 5 for which the critical value is 9.488. The $\chi^{2}$ value for the d$_{c}$ = 16.6 nm condition is 7.85, which means that the null hypothesis is accepted, and the orientations are deemed random. Note, due to the sparse data set for the d$_{c}$ = 16.6 nm starting condition, the Yates' correction for continuity \cite{YatesCorrection} is applied here.

Figure \ref{fig:Fig4}(b) shows the grain orientations for the d$_{c}$ = 5.2 nm cells (20 Grain Voronoi). The number of bins is 20 for which the critical value is 30.144. The $\chi^{2}$ value between the theoretical and observed distributions for this case is 8.8, so once again the null hypothesis is accepted. 138 bins are employed when considering the sheared simulations' grain orientation distribution in Figure \ref{fig:Fig4}(c). The critical value in this case is 113.605, whilst the $\chi^{2}$ value is 196.830. As such, the null hypothesis is rejected and we conclude that, as shown in Figure \ref{fig:Fig3}, the starting orientations of the sheared cells are not random.

\subsection{Simulated Evolution of Microstructure Upon Irradiation}

\begin{figure*}[t]
\includegraphics[width=\linewidth]{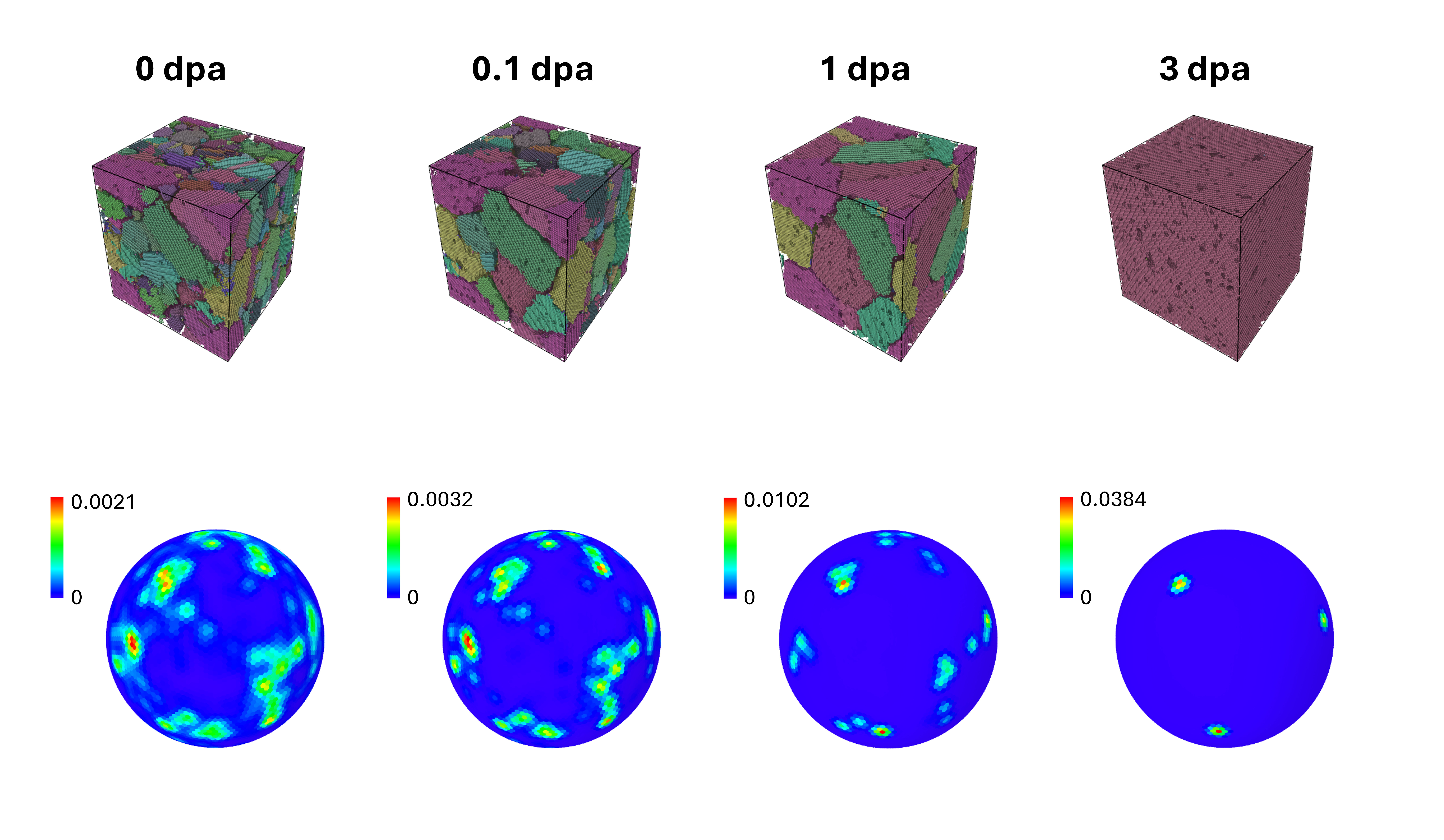}
\caption{Illustration of grain growth for a single case in a \textit{sheared} cell. Grain coarsening shown visually in the top row with increased dose, and through corresponding pole figures in the bottom row. This behaviour present in all initially nanocrystalline simulations with increasing dose.}
\label{fig:Fig5}
\end{figure*}

\begin{figure}
\begin{subfigure}{0.43\textwidth}
\includegraphics[width=\linewidth]{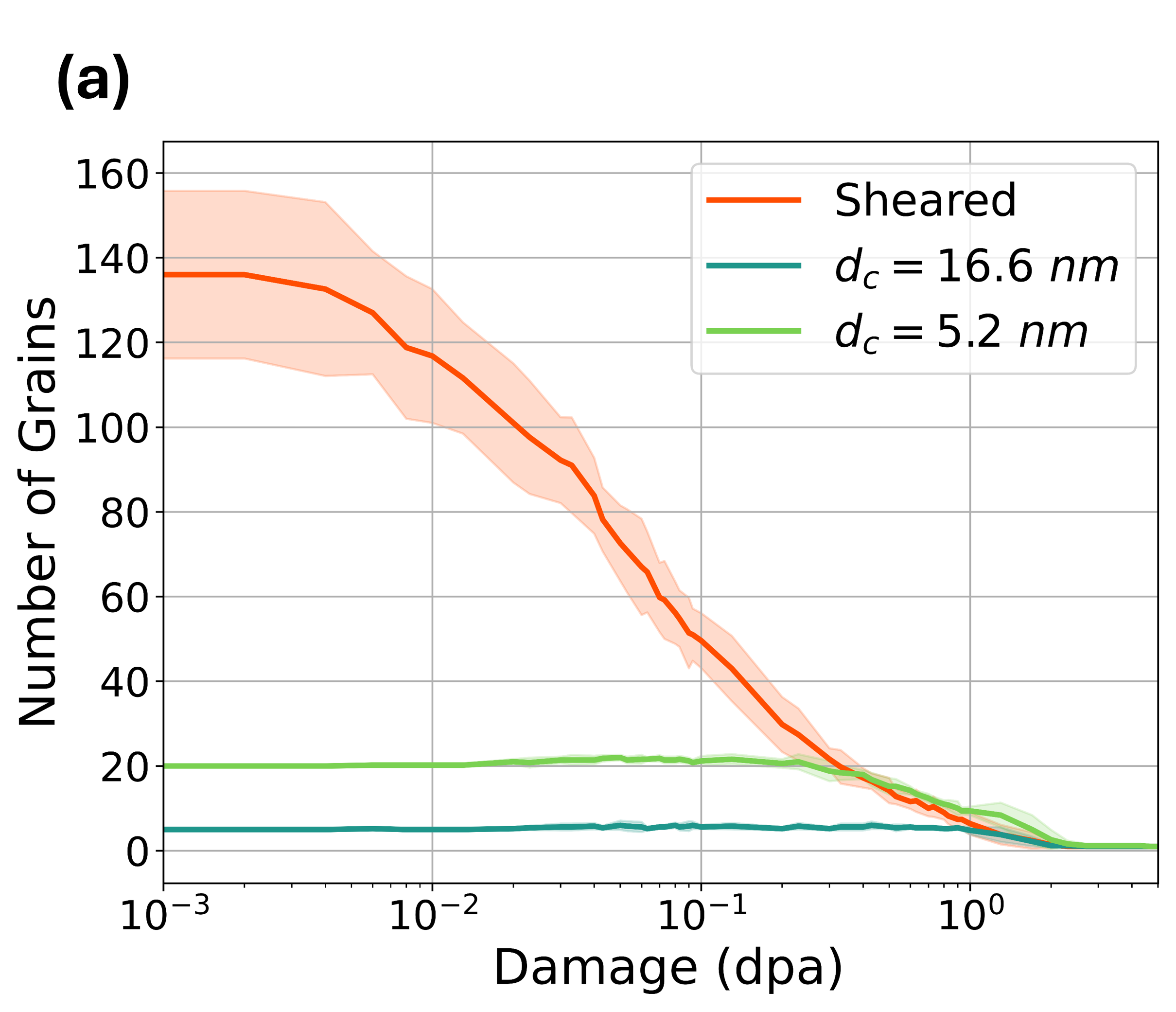}
\label{fig:Fig6a}
\end{subfigure}

\begin{subfigure}{0.43\textwidth}
\includegraphics[width=\linewidth]{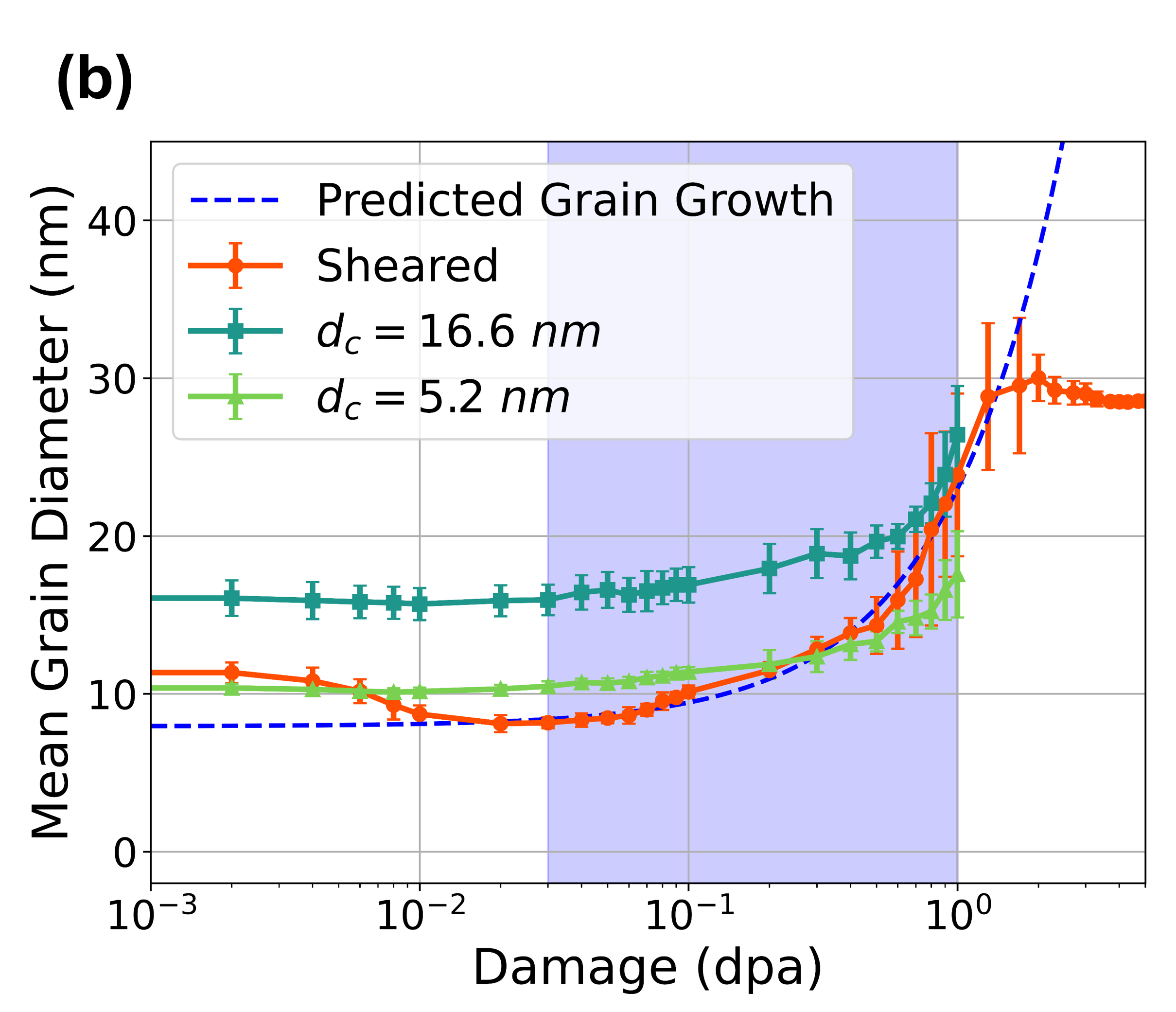}
\label{fig:Fig6b}
\end{subfigure}

\begin{subfigure}{0.43\textwidth}
\includegraphics[width=\linewidth]{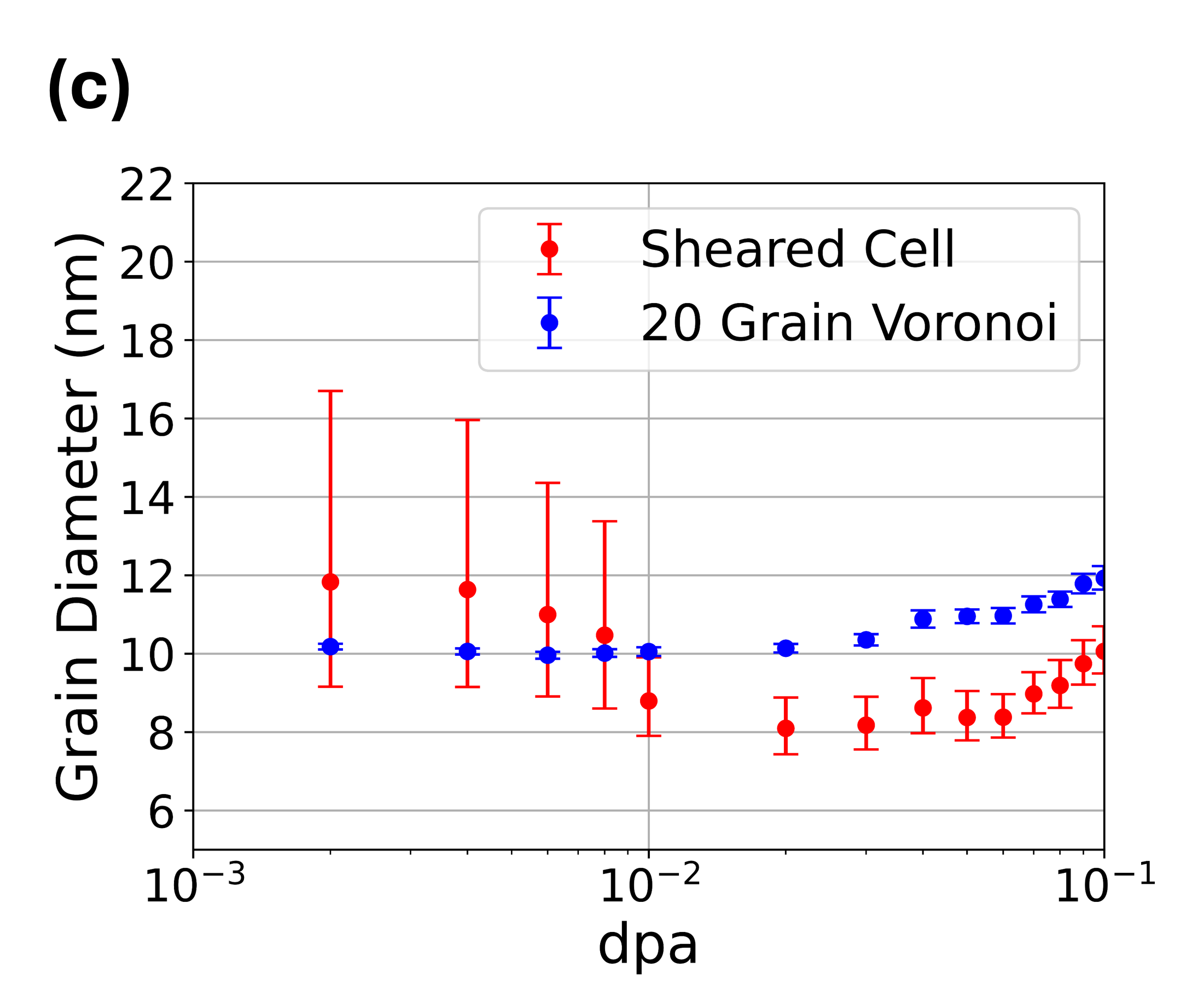}
\label{fig:Fig6c}
\end{subfigure}
\caption{Grain size evolution with increased dose for the initially nanocrystalline cells: (a) Grain number as a function of dose for the Voronoi-generated and \textit{sheared} cells. The smallest permissible grain is taken as 100 atoms, (b) Grain diameter as a function of dose for the initially nanocrystalline cells based on Williamson-Hall analysis, (c) Single instance of grain diameter for a sheared cell and 20 grain Voronoi created cell using Williamson-Hall.} \label{fig:Fig6}
\end{figure}

Figure \ref{fig:Fig5} shows a key result of this work. All of the nanocrystalline cells experience a grain growth with increased irradiation, ultimately reducing to a single crystal form. This can also be seen in Figure \ref{fig:Fig6}(a) where the average number of grains for the five distinct simulations is plotted as a function of dose for both the sheared and the Voronoi-created cells. For this analysis, a grain was taken to be at least 100 atoms.

It is evident from Figure \ref{fig:Fig6}(a) that the sheared cells start with far more grains than the Voronoi-created cells with the reduction in grain number begining at different stages for the different simulations. At $\sim$ 0.003 dpa, the average grain number in the sheared cells begins to decrease as irradiation induced grain growth occurs. This process continues until a single grain remains after $\sim$ 2 dpa. However, the grain number does not begin to decrease for the 20 grain Voronoi (d$_{c}$ = 5.2 nm) and 5 grain Voronoi cells (d$_{c}$ = 16.6 nm) until $\sim$ 0.25 dpa and $\sim$ 1 dpa, respectively. This suggests that there is a relationship between starting grain size and the point at which the grains grow due to irradiation. If the volume exchange rate per dpa between grains is constant, then smaller grains would be removed sooner than large grains, which is seemingly supported by Figure \ref{fig:Fig6}(a).

To investigate whether the simulation cell size influences the upper limit of grain growth, a nanocrystalline iron needle, with aspect ratio  $\sim$ 1:9, was irradiated up to 2.5 dpa, shown in Appendix D. Appendix D shows the results of a collision cascade simulation carried out to 2.5 dpa, on a 100 grain cell with $\sim$ 5 million atoms. As per Appendix D, we cannot rule out the existence of a grain size limit after which the irradiation-induced grain growth will stop. Nonetheless, this work aims to gain mechanistic insight into the evolution of microstructure under irradiation, so the size effect is not considered further.

Irradiation induced grain growth has been observed in previous MD simulations of nanocrytalline materials \cite{LeoMaPRM, LevoCascades}. Ma \textit{et al.} \cite{LeoMaPRM} showed an increase in grain size with irradiation when considering NC tungsten, a BCC metal. It was found that the largest increase in grain diameter occurred in simulations of smallest initial grains. This is directly comparable to Figure \ref{fig:Fig6} whereby, since all simulations become single crystalline at roughly the same dose, the difference in grain size is the largest for the sheared case. Levo \textit{et al.} \cite{LevoCascades} also showed similar evolution in nanocrystalline nickel, with the NC cells becoming single crystalline with increasing dose. Thus our observations of irradiation-induced grain growth are consistent with the existing computational literature.

By utilising Williamson-Hall analysis on the MD data, the mean grain diameter as a function of dose for the initially nanocrystalline cells is plotted in Figure \ref{fig:Fig6}(b). This was calculated by generating line profiles at various damage levels and applying Williamson-Hall analysis, averaged over five distinct cells, consistent with experimental methods. Note, the Voronoi data ends at 1 dpa, while sheared data extends to 5 dpa. Notably, peak splitting occurred in Voronoi cells beyond $\sim$ 1 dpa, suggesting a transition toward a single crystal with potential tetragonal distortion, making Williamson-Hall analysis unsuitable beyond this point.

Whilst this is a different method of calculating the grain size, it is evident that the grain growth process present in Figure \ref{fig:Fig6}(a) is confirmed, with all cells showing an increase in mean grain diameter with dose. This analysis was particularly useful for comparison with the experimental data. The sheared cell data was used to predict grain growth in experimental data with a fitted function, shown with the dashed blue line. This was derived between 0.03 dpa and 1 dpa, as shown with the light blue box. This damage range was chosen as the data below 0.03 dpa is quite noisy with large errors due to many small grains being present. Figure \ref{fig:Fig6}(c) shows a single instance of a sheared cell against a 20 grain Voronoi-created cell. Evidently, at low damage, the errors in grain diameter are large compared to the Voronoi case and these errors greatly reduce at $\sim$ 0.03 dpa. An in depth analysis can be found in the Discussion section, below.

\begin{figure}
\begin{subfigure}{0.43\textwidth}
\includegraphics[width=\linewidth]{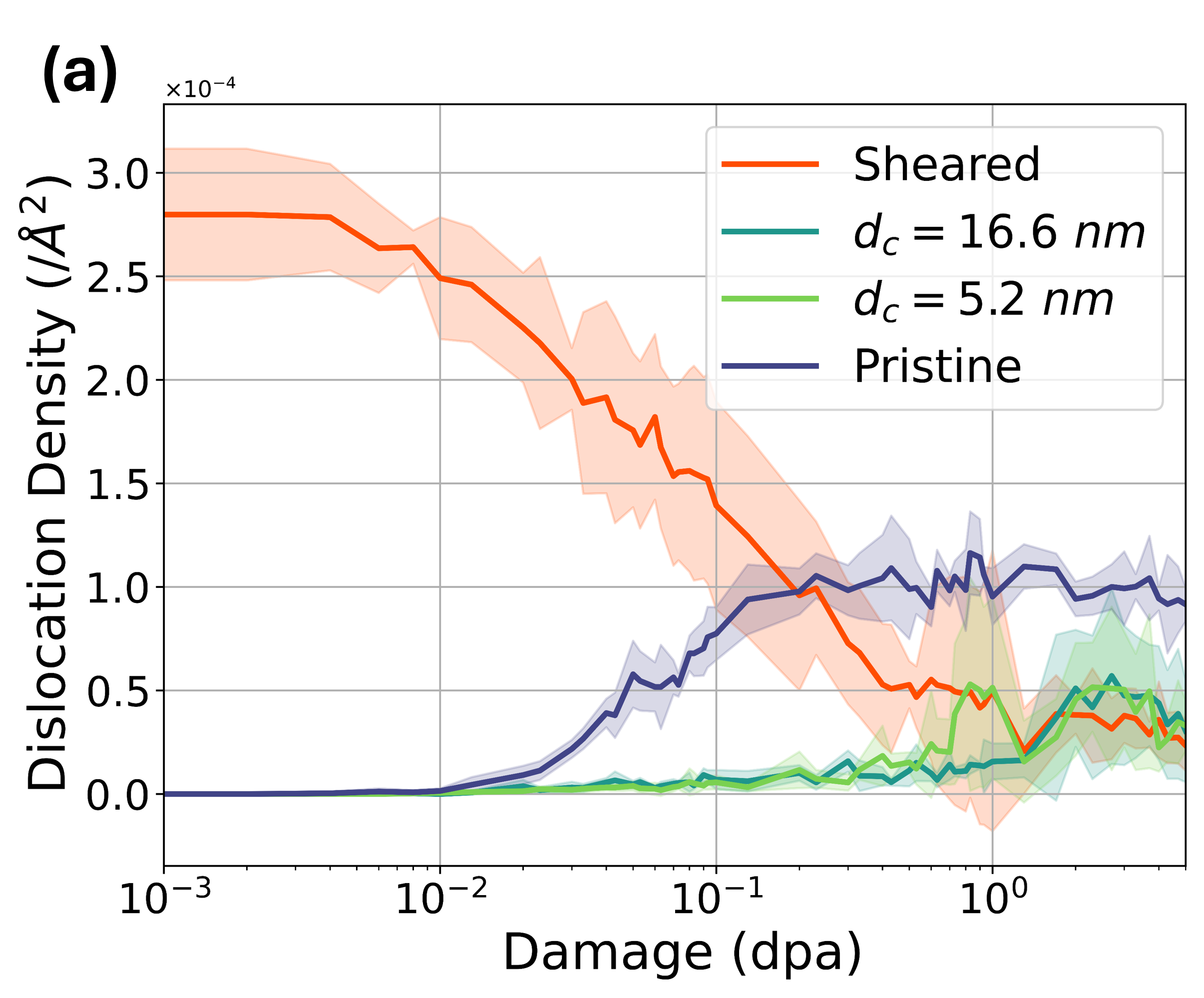}
\label{fig:Fig7a}
\end{subfigure}

\begin{subfigure}{0.43\textwidth}
\includegraphics[width=\linewidth]{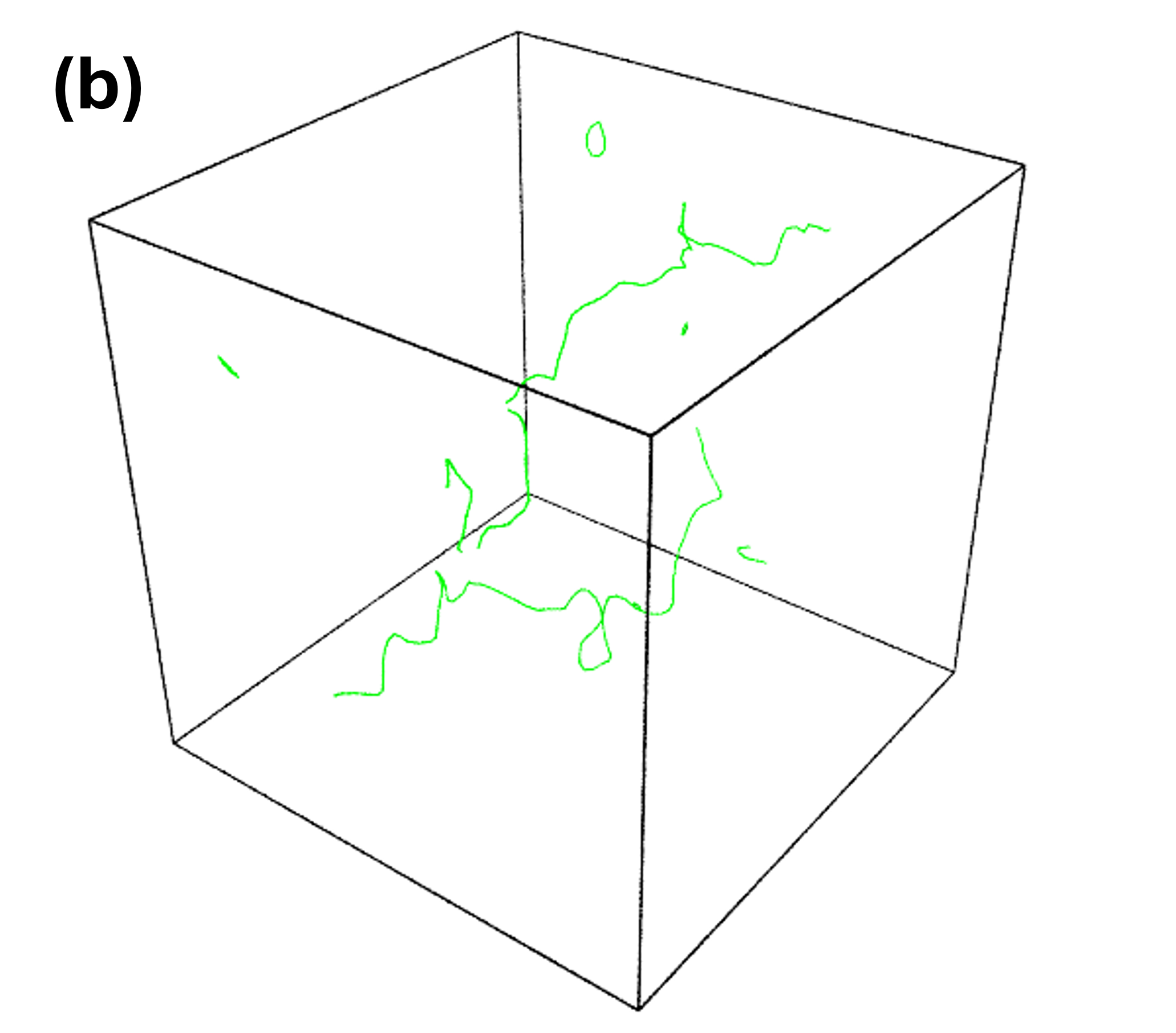}
\label{fig:Fig7b}
\end{subfigure}

\begin{subfigure}{0.43\textwidth}
\includegraphics[width=\linewidth]{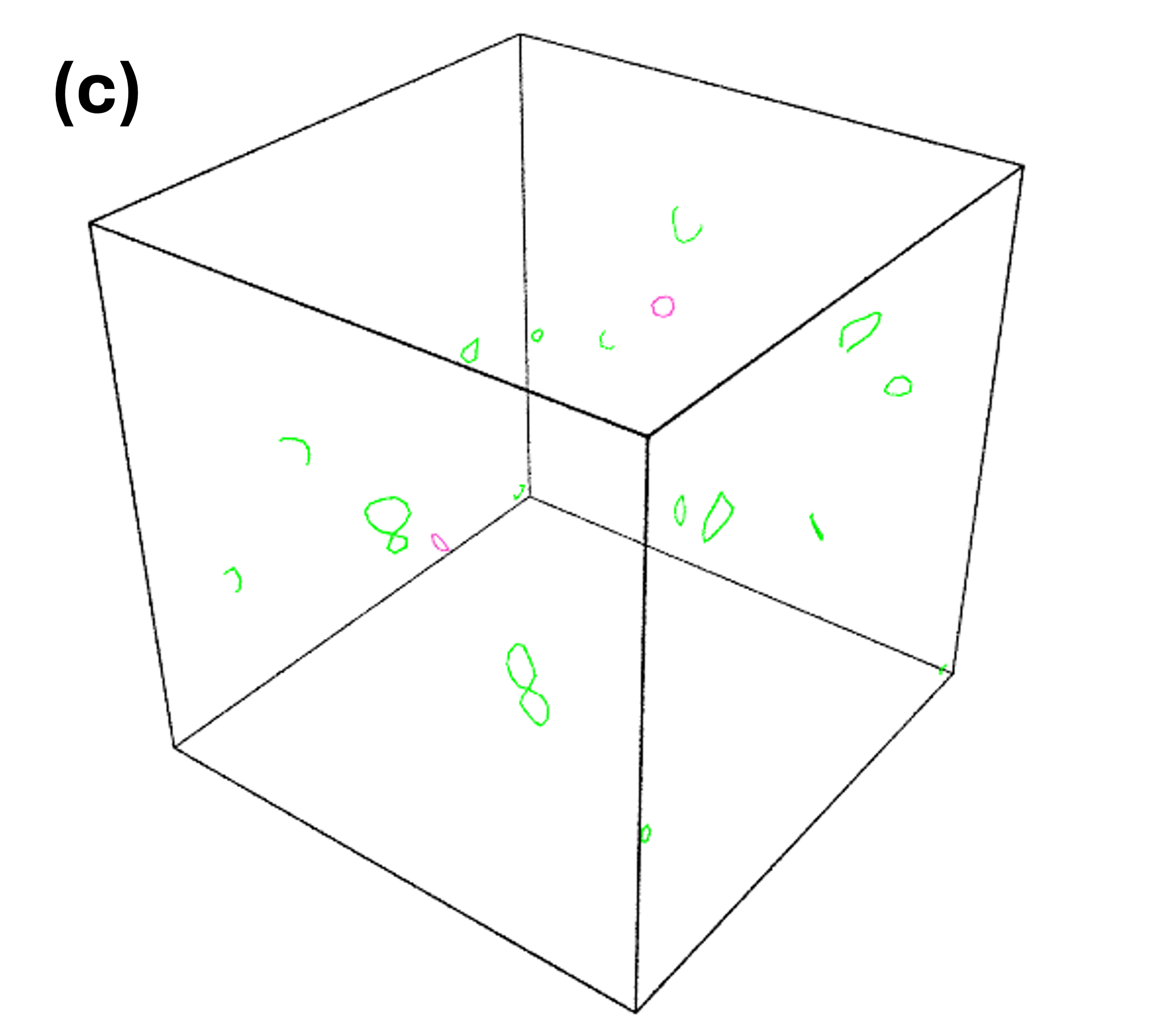}
\label{fig:Fig7c}
\end{subfigure}

\caption{Comparison of dislocation density between simulations; (a) Dislocation density as a function of dose, (b) Visualisation of dislocations for an initially pristine cell at 5 dpa, (c) Visulisation of dislocations for an initially nanocrystalline cell at 5 dpa.} \label{fig:Fig7}
\end{figure}

Figure \ref{fig:Fig7}(a) depicts the dislocation density in the nanocrystalline cells as a function of dose, and compares these to pristine, single crystal cells. When computing the dislocation density, atoms with  $d<4$ {\AA} from a grain boundary are removed so that the in-grain dislocation density is displayed. Initially, at 0 dpa, there are no dislocations in the Voronoi or pristine cells, whilst the sheared cells have a dislocation density of $\sim2.8 \times 10^{-4}/$\AA$^2$ due to severe plastic deformation applied during their creation.

For the initially pristine cells, the dislocation density begins to increase at $\sim$ 0.01 dpa, increasing to $\sim1.0 \times 10^{-4}/$\AA$^2$ at $\sim$ 0.2 dpa and then remaining roughly constant until 5 dpa. Similarly, for the initially 20 grain Voronoi (d$_{c}$ = 5.2 nm) and 5 grain Voronoi cells (d$_{c}$ = 16.6 nm), the dislocation density begins to increase at $\sim$ 0.01 dpa however, much slower than the rate at which the initially pristine samples increase. Whilst the initially pristine cells increase to $\sim1.0 \times 10^{-4}/$\AA$^2$ at $\sim$ 0.2 dpa, the Voronoi cells show around $\sim1.0 \times 10^{-5}/$\AA$^2$ at the same dose. It is not until $\sim$ 2.0 dpa that the dislocation density saturates for both the initially 5 grain and 20 grain Voronoi cells, at a value of $\sim 5 \times 10^{-5}/$\AA$^2$, half of that observed in the initially pristine cells.

The dislocation density in the sheared cells follows a different evolution as it does not start from zero. From 0 dpa to $\sim$ 0.003 dpa, the dislocation density remains constant at $\sim2.8 \times 10^{-4}/$\AA$^2$. After this dose, the dislocation density steadily decreases until saturating at $\sim$ 2.0 dpa, the same point as the Voronoi cells, at a value of $\sim3.0 \times 10^{-5}/$\AA$^2$. Overall, Figure \ref{fig:Fig7}(a) clearly shows that starting with initially nanocrystalline samples results in a reduced dislocation density at high irradiation doses compared to initially pristine samples, notwithstanding the fact that the nanocrystalline samples become single crystalline with irradiation. The dislocation density of the initially nanocrystalline cells is surprisingly similar at 5 dpa, irrespective of whether they were created through Voronoi tessellation or shear.

Song \textit{et al.} \cite{SONG2024114144} showed experimentally a decrease in dislocation density with irradiation for nanocrystalline Eurofer97, whilst the same study also showed an increase in dislocation density with irradiation for coarse-grained material. It appears that the presence of grain boundaries, for the Voronoi and sheared cases, acts as a sink for irradiation-induced defects as reported experimentally \cite{ELATWANI2019118, ROSE1997119, Dey}, and in line with the simulation data in Ma \textit{et al.} \cite{LeoMaPRM}. However, this does not explain why the dislocation density for the initially nanocrystalline cells remains much lower than that of the pristine cells even when all nanocrystalline boxes have become single crystalline at doses $>$ 2.5 dpa. 

Further analysis was carried out to explore why the initially NC cells exhibit a lower dislocation density than the pristine cells, even though the NC cells become single crystalline at high damage levels. Figures \ref{fig:Fig7}(b) and \ref{fig:Fig7}(c) show the dislocation lines, obtained through DXA, for a single instance of a pristine cell and initially nanocrystalline cell, respectively, at 5 dpa. Here, green lines represent dislocations with Burgers vector $\mathbf{b}=\frac{1}{2}\langle 111 \rangle$, and pink lines represent $\mathbf{b}=\langle100\rangle$ dislocations. It is immediately obvious that the nature of dislocations in the two cells is very different. In Figure \ref{fig:Fig7}(b), the dislocations have coalesced to form long lines that span the entirety of the cell. However, in Figure \ref{fig:Fig7}(c), many small dislocation loops are present and there does not appear to be any coalescence into a large dislocation network.

\begin{figure}
\begin{subfigure}{0.48\textwidth}
\includegraphics[width=\linewidth]{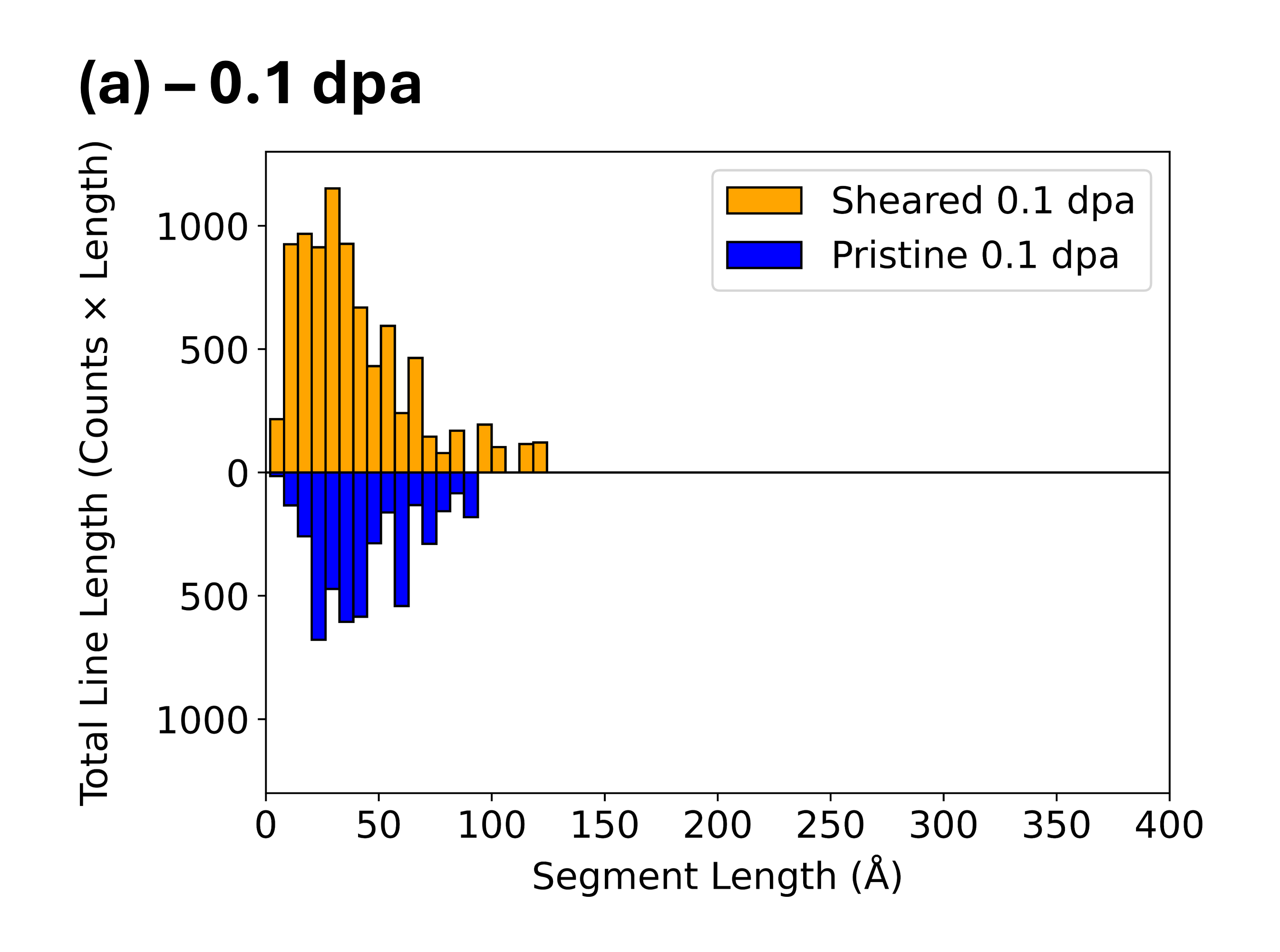}
\label{fig:Fig8a}
\end{subfigure}

\begin{subfigure}{0.48\textwidth}
\includegraphics[width=\linewidth]{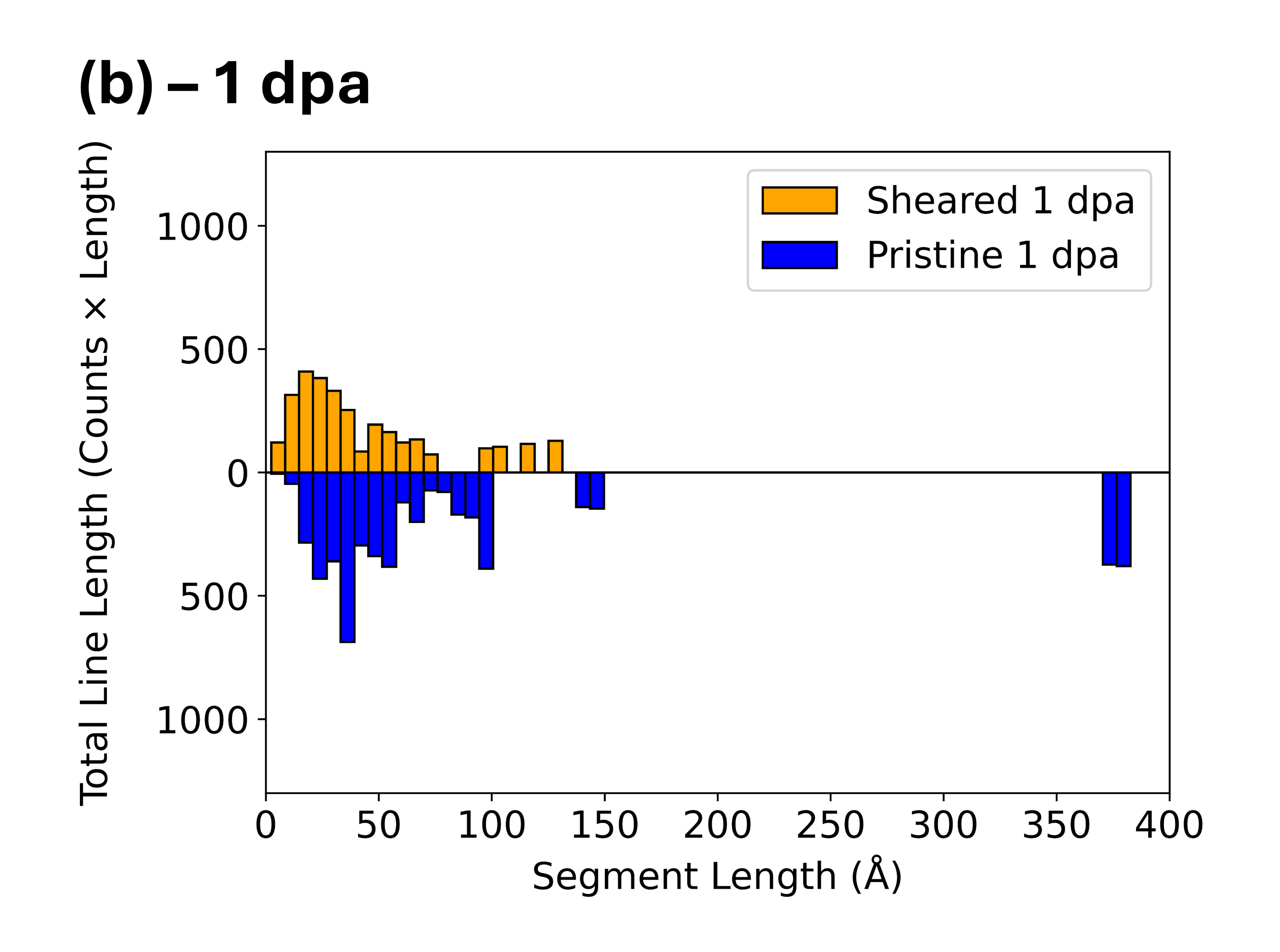}
\label{fig:Fig8b}
\end{subfigure}

\begin{subfigure}{0.48\textwidth}
\includegraphics[width=\linewidth]{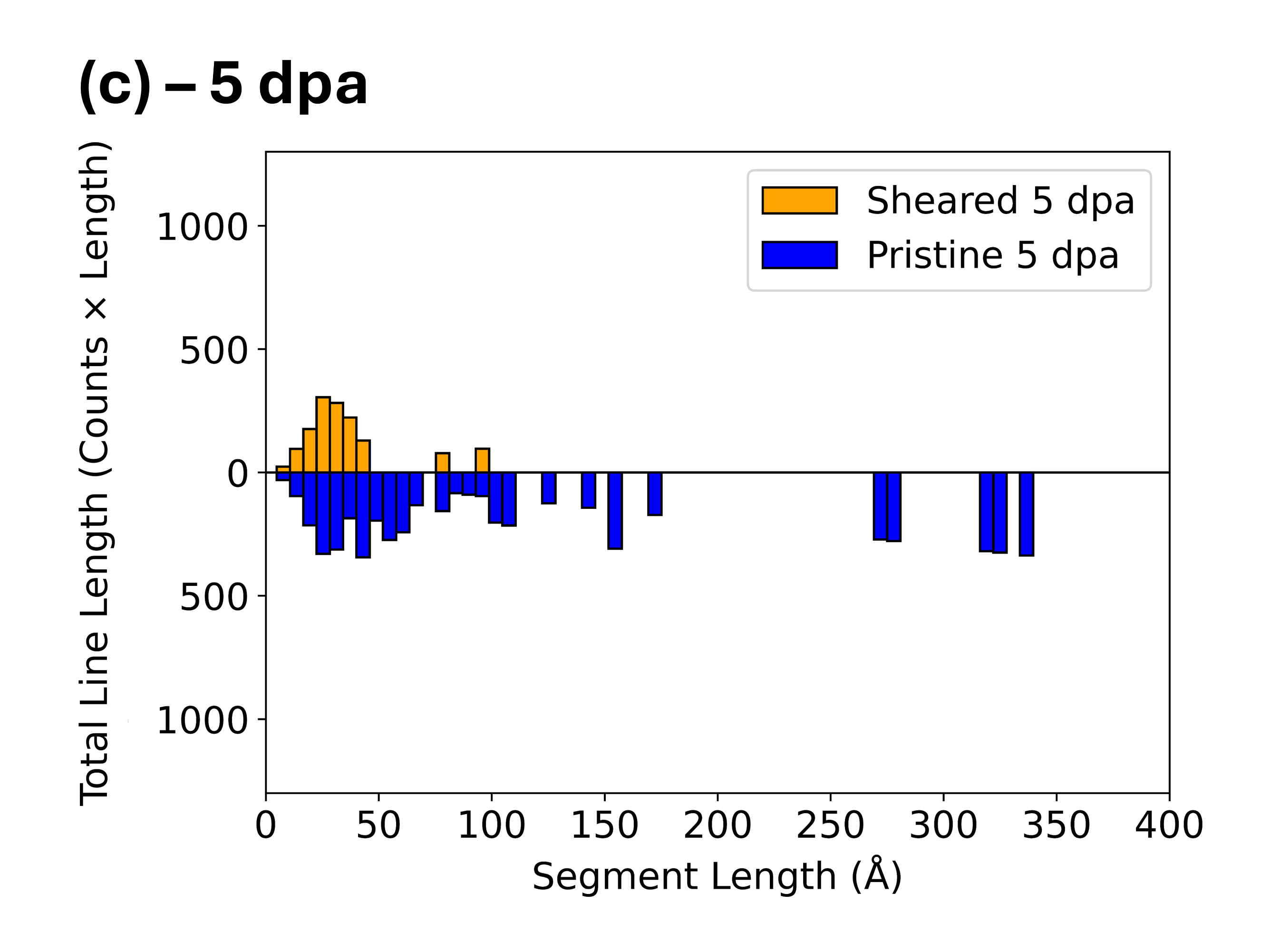}
\label{fig:Fig8c}
\end{subfigure}
\caption{Histogram of dislocation segment lengths against total line length for the pristine and sheared simulations; (a) 0.1 dpa, (b) 1 dpa, (c) 5 dpa.}
\label{fig:Fig8}
\end{figure}

Figure \ref{fig:Fig8} shows histograms of dislocation segment lengths versus total line lengths across different damage levels, comparing sheared and pristine cells (averaged over five simulations each). At 0.1 dpa (Figure \ref{fig:Fig8}(a)), both cell types exhibit short segments below 150 \AA. At 1 dpa (Figure \ref{fig:Fig8}(b)), segment lengths in sheared cells remain similar, while pristine cells show a clear increase, with some reaching $\sim$ 380 \AA. By 5 dpa (Figure \ref{fig:Fig8}(c)) a clear difference emerges: sheared cells retain short segments, while pristine cells display both short and large segments exceeding 150 \AA, consistent with visual trends in Figures \ref{fig:Fig7}(b) and \ref{fig:Fig7}(c).

Our hypothesis for this behaviour is as follows: In the pristine cells, there are no grain boundaries to limit the growth of irradiation-induced dislocations. These dislocations are free to move and coalesce to form large networks within the cell. With higher irradiation, these large networks persist and a fossil of this large dislocation network is left over that would be too energetically costly to remove. Conversely, the presence of grain boundaries in the initially NC cells constrains the volume in which dislocations can form, and the grain boundaries can even remove some irradiation defects by acting as a sink. As such, large dislocation networks are unable to form and coalesce and only small dislocation loops persist as shown in Figures \ref{fig:Fig7}(c) and \ref{fig:Fig8}. This suggests that the starting microstructure of iron is important for prediction of its evolution under irradiation, even if everything becomes single crystalline at high damage levels.

\begin{figure}
\begin{subfigure}{0.47\textwidth}
\includegraphics[width=\linewidth]{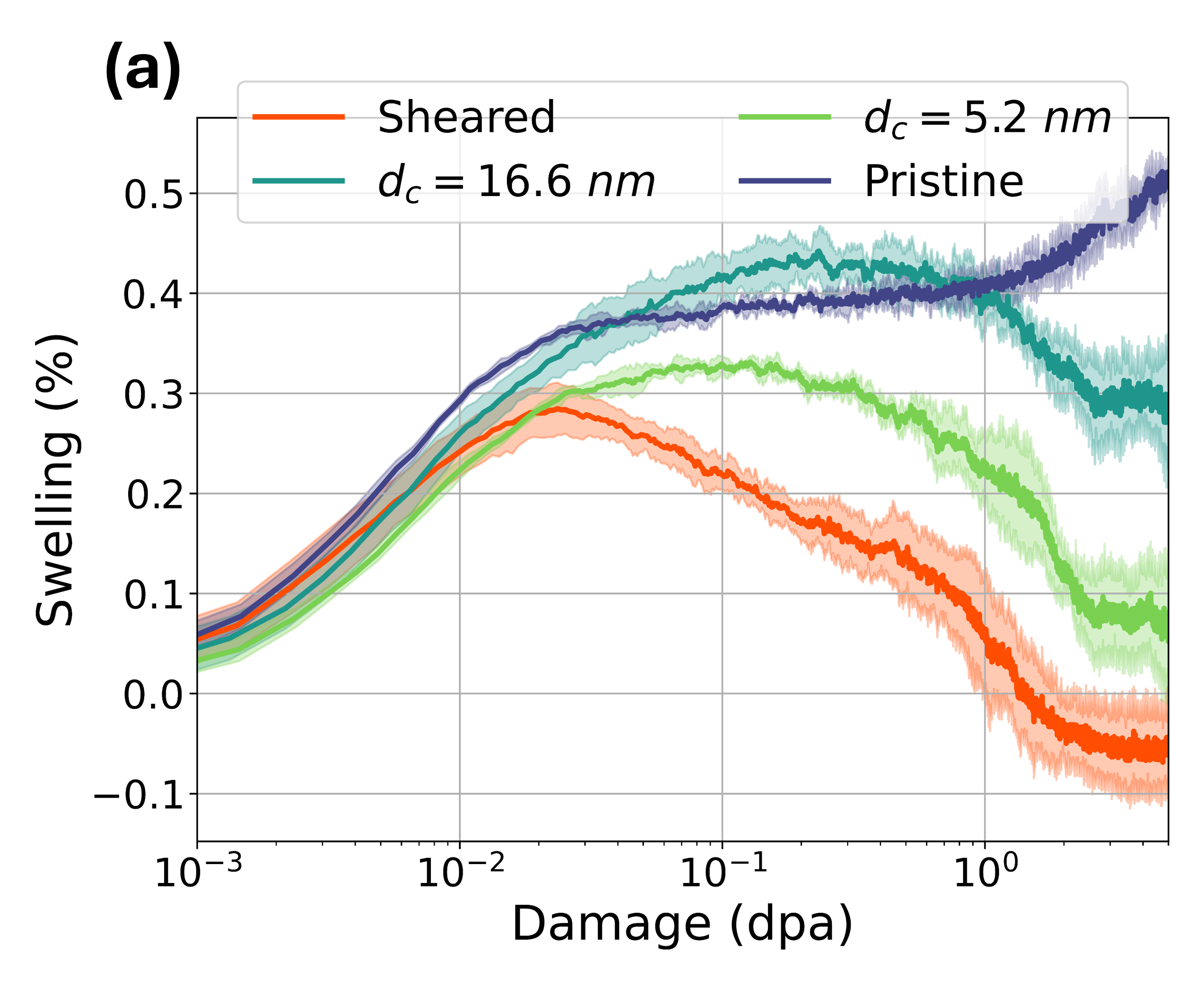}
\label{fig:Fig9a}
\end{subfigure}
\begin{subfigure}{0.47\textwidth}
\includegraphics[width=\linewidth]{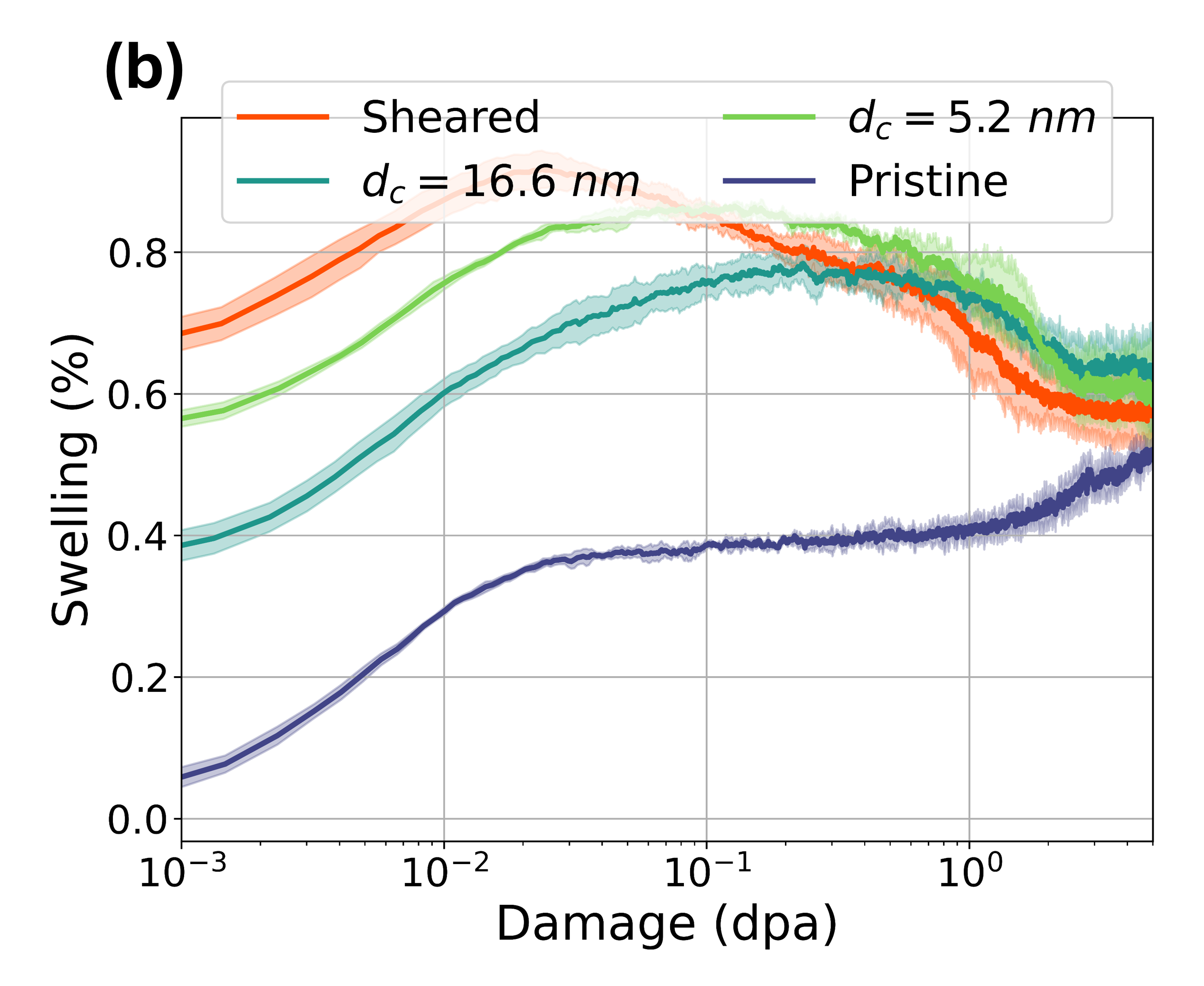}
\label{fig:Fig9b}
\end{subfigure}
\begin{subfigure}{0.47\textwidth}
\includegraphics[width=\linewidth]{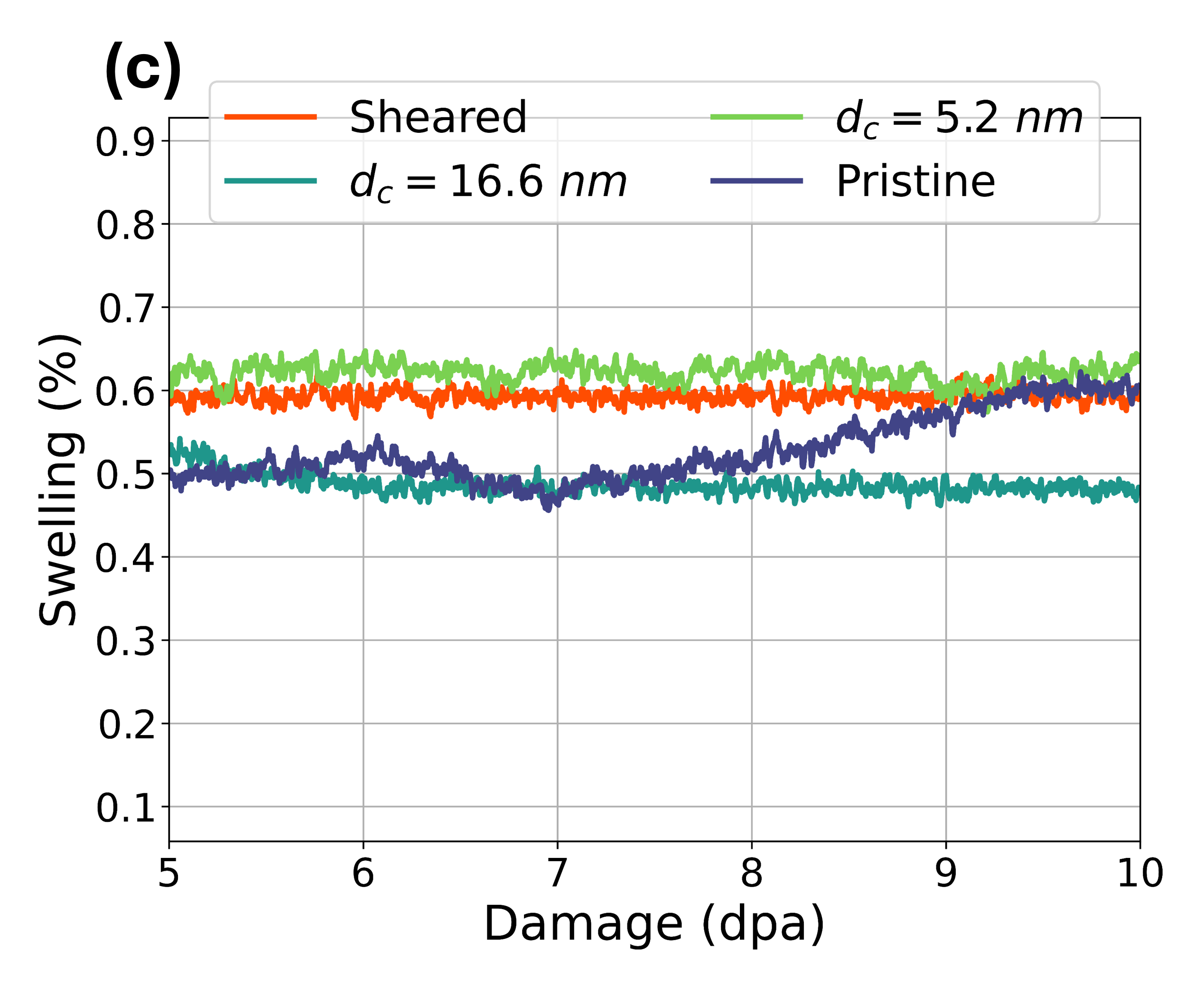}
\label{fig:Fig9c}
\end{subfigure}
\caption{Comparison of volumetric swelling between simulations; (a) Swelling percentage benchmarked against each cell at 0 dpa, (b) Swelling percentage benchmarked against the pristine cell at 0 dpa, and (c) Swelling for one distinct simulation cell of each simulation condition from 5 to 10 dpa benchmarked against the pristine cell at 0 dpa.} \label{fig:Fig9}
\end{figure}

The swelling behaviour of the material with dose also differs between the initially pristine, Voronoi, and sheared cells, as shown in Figure\ref{fig:Fig9}(a). Swelling is calculated relative to a reference sample at 0 dpa for each individual simulation from the cell box side lengths. Up to $\sim$ 0.01 dpa, the swelling behaviour of all simulations is very similar, increasing roughly at the same rate with dose. After this point, the initially pristine samples continue increasing steadily, finishing at $\sim$ 0.52\% swelling at 5 dpa. 

The nanocrystalline samples do not follow this trajectory. All of these samples show an increase in irradiation induced swelling, followed by a decrease. The cells with initially d$_{c}$ = 16.6 nm grains increase to a maximum swelling of 0.42\% at 0.2 dpa whilst the cells with initially d$_{c}$ = 5.2 nm grains increase to a maximum of 0.32\% at 0.1 dpa. With dose, the d$_{c}$ = 16.6 nm grain cells decrease from this maximum point, before plateauing at $\sim$ 0.3\% swelling after 2 dpa. Likewise, the d$_{c}$ = 5.2 nm grain cells decrease from the maximum point to $\sim$ 0.09\% swelling at around 3 dpa.

Similarly, the irradiation induced swelling increases in the sheared cells to a maximum point of $\sim$ 0.28\% at 0.02 dpa before decreasing. Note that the maximum swelling in the sheared cells is lower than that for any other simulation type, and occurs at a much lower dose. Interestingly, for the case of the sheared cells, the swelling becomes negative when compared to the 0 dpa case, reducing to and plateauing at $\sim$ - 0.06\% at 2 dpa.

Similar behaviour has previously been observed in CRA simulations on nanocrystalline Tungsten. Ma \textit{et al.} \cite{LeoMaPRM} showed that materials with smaller initial grain size experienced less swelling than coarser-grained and single crystal materials, and this is evident in this work, shown in Figure \ref{fig:Fig9}(a). The presence of vacancies and voids within the material directly relates to the swelling and the interaction of these defects with the grain boundaries results in a lower irradiation induced swelling compared to initially pristine, single crystalline samples \cite{LeoMaPRM}. In fact, Ma \textit{et al.} \cite{LeoMaPRM} suggests that materials with a more damaged initial microstructure experience less irradiation induced swelling with dose, as confirmed here.

Figure \ref{fig:Fig9}(b) also shows the swelling when it is referenced to an initially pristine cell, that is the $V_{0}$ value is always taken as that for an initially pristine cell. In Figure \ref{fig:Fig9}(b), the sheared cells initially show the largest swelling of $\sim$ 0.7\% compared to the pristine case at 0.001 dpa, whilst the d$_{c}$ = 5.2 nm and the d$_{c}$ = 16.6 nm cells show $\sim$ 0.55\% and $\sim$ 0.4\%, respectively. Interestingly, at 5 dpa, all the simulations have roughly the same swelling of $\sim$ 0.6\%, with the sheared cell being slightly lower. This is a significant result as it suggests that after a certain damage level, the material forgets its initial nanocrystalline configuration and ultimately saturates at a certain swelling value irrespective of whether the cells were created with Voronoi tessellation or shear. 

It also appears that the pristine case is rising to this value and would potentially saturate at a similar point to the initially nanocrystalline cells. To test this hypothesis, a single instance of all different cell types was run to 10 dpa, presented in Figure \ref{fig:Fig9}(c). This data shows that the swelling in the pristine cell continues to rise until $\sim$ 9.4 dpa, when it is very similar to the swelling in the sheared and d$_{c}$ = 5.2 nm cells, after which is plateaus at near identical values to the initially nanocrystalline cells. The data appears to suggest that there is a late-blooming phase in the swelling evolution of pristine iron, evident at high doses. It appears that irrespective of the material's starting configuration, the swelling will eventually saturate at the same point for high doses above 9 dpa. Note that the swelling in the d$_{c}$ = 16.6 nm cell decreases slightly, and plateaus at a lower swelling value and this is attributed to sampling as only one instance of each cell type was run to 10 dpa.

Experimentally, many studies report a greater resistance of fine grained materials against irradiation induced swelling when compared to coarse-grained materials for both ferritic alloys \cite{GIGAX2017395, SONG2014285} and other materials \cite{SRINIVASAN2020621, SUN_SS304L}. Song \textit{et al.} \cite{SONG2014285} found that the swelling rate of ultra fine-grained T91 was three times lower than that of coarse-grained T91, which was attributed to the presence of many defect sinks in the ultra fine-grained material.

\begin{figure}
\begin{subfigure}{0.5\textwidth}
\includegraphics[width=\linewidth]{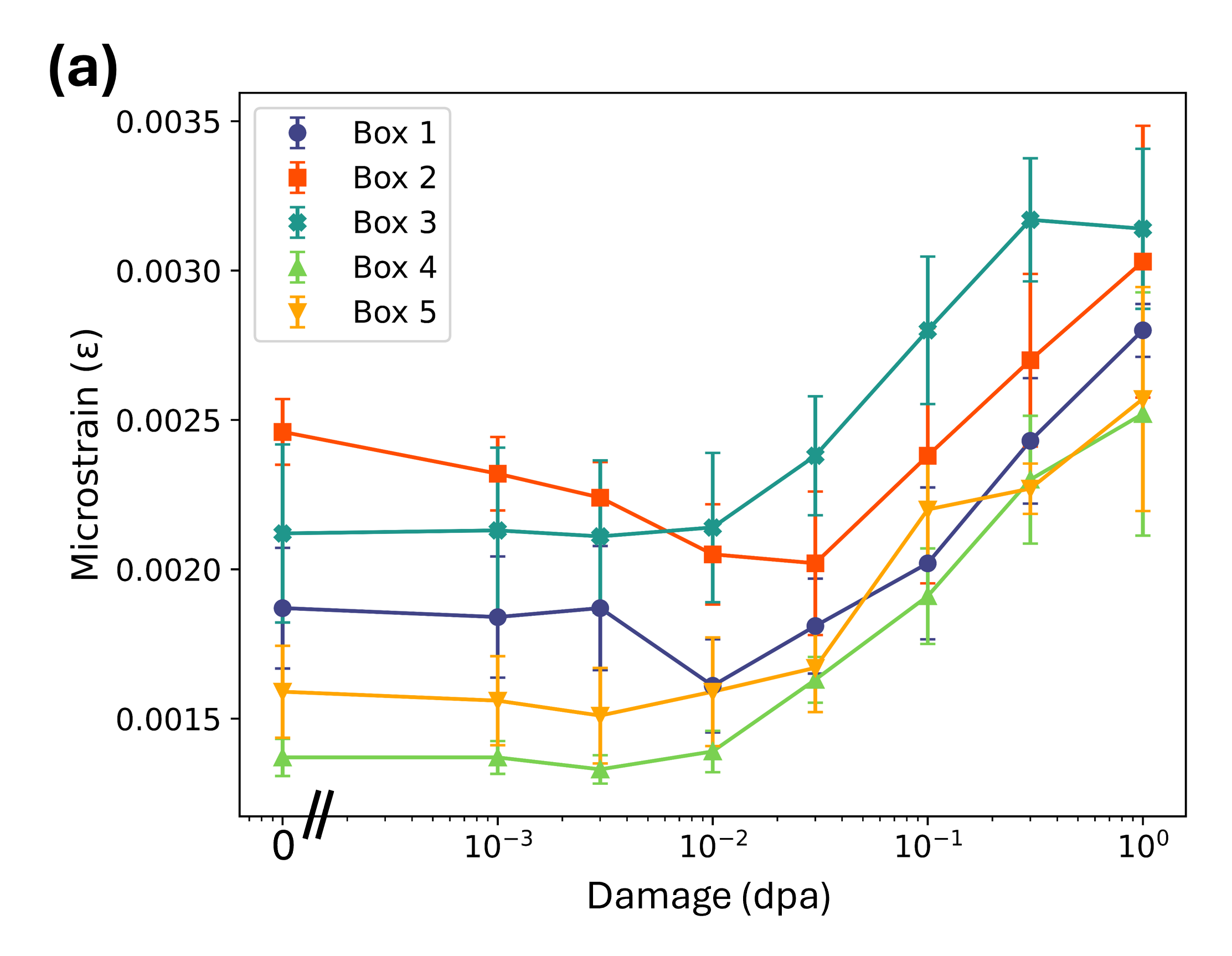}
\phantomsubcaption\label{fig:Fig10a}
\end{subfigure}

\begin{subfigure}{0.5\textwidth}
\includegraphics[width=\linewidth]{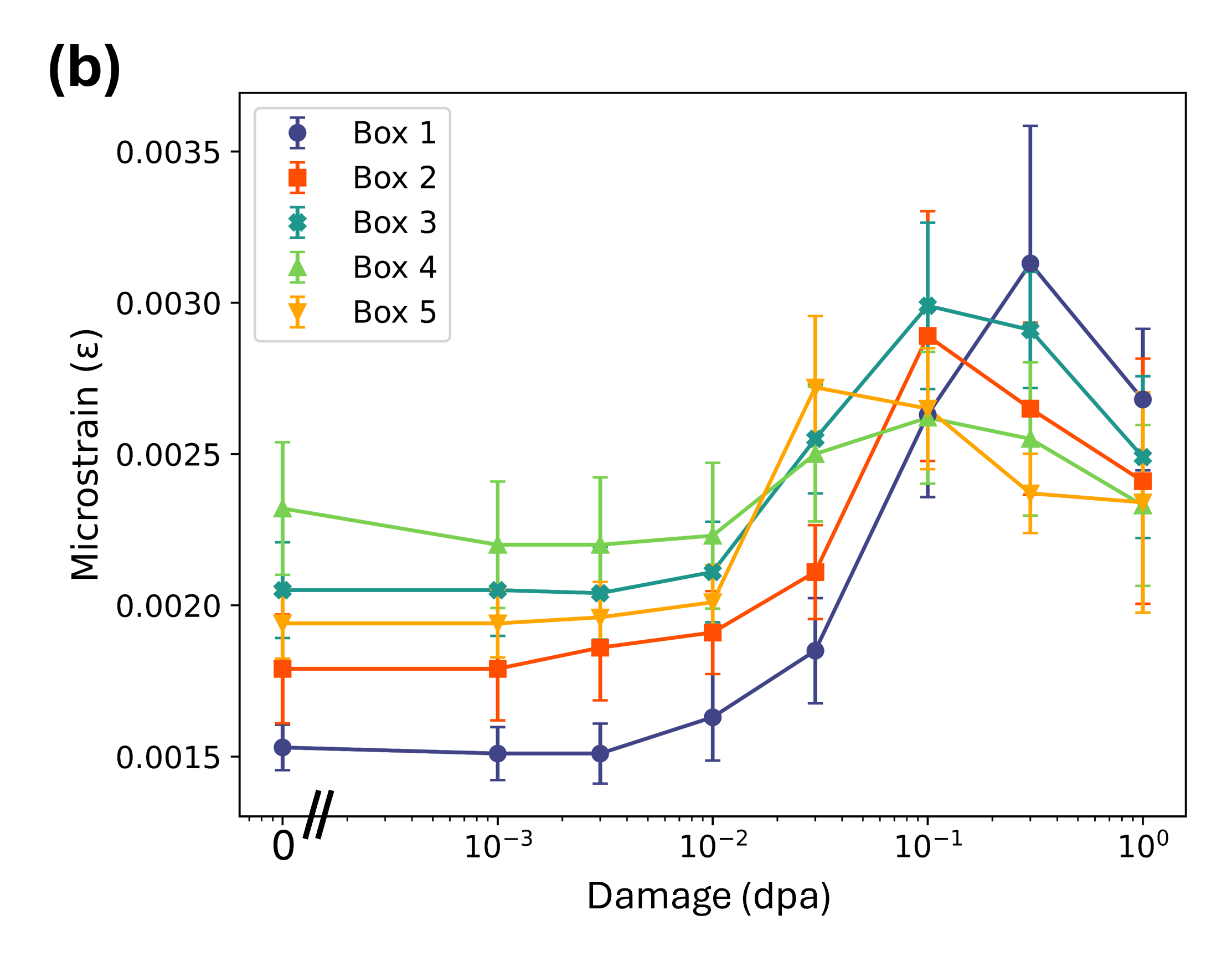}
\phantomsubcaption\label{fig:Fig10b}
\end{subfigure}

\begin{subfigure}{0.5\textwidth}
\includegraphics[width=\linewidth]{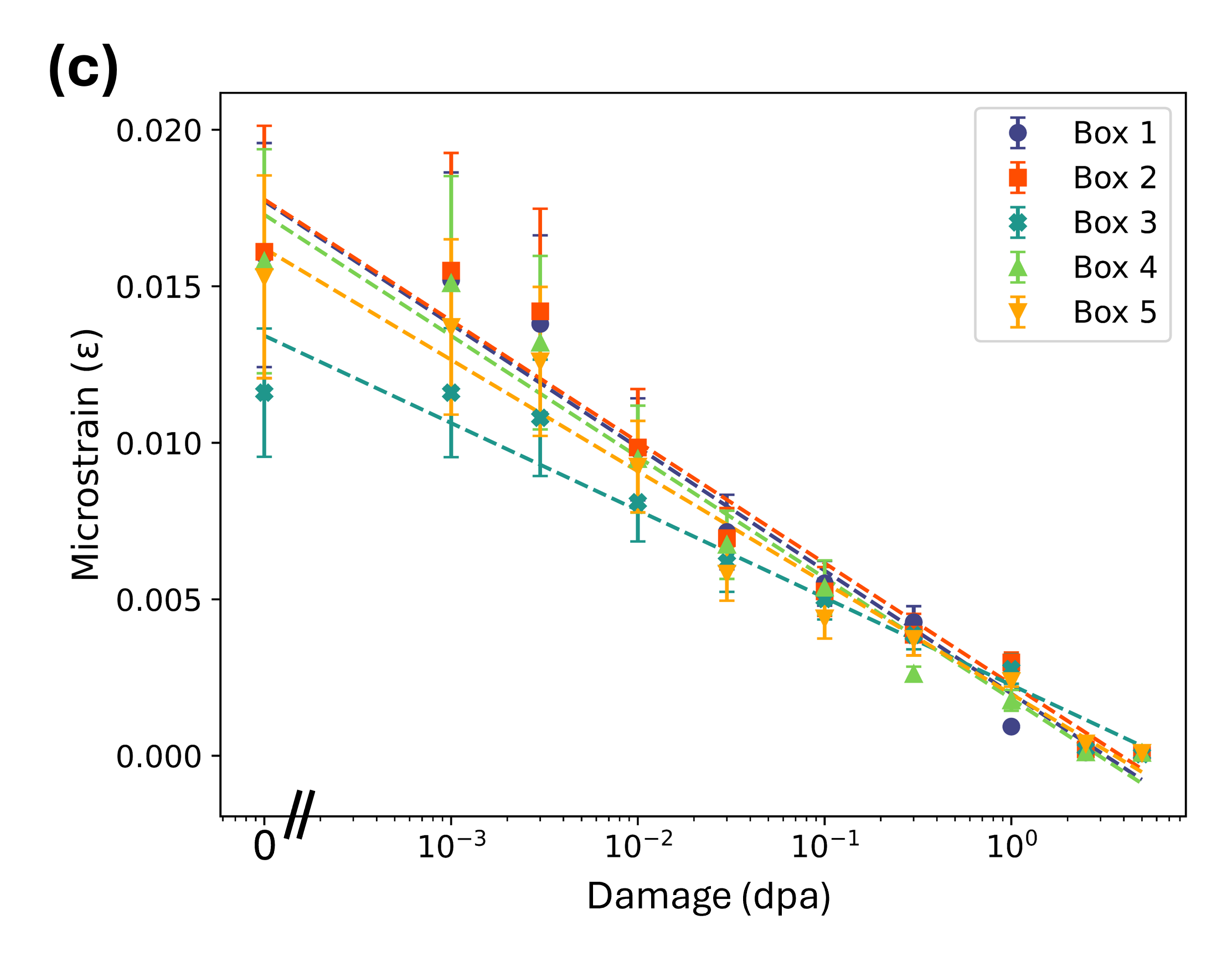}
\phantomsubcaption\label{fig:Fig10c}
\end{subfigure}

\caption{Microstrain obtained through Williamson-Hall analysis on line profiles for the MD cells calculated through the Debye scattering equation; (a) d$_{c}$ = 16.6 nm cells, (b) $_{c}$ = 5.2 nm cells, (c) Sheared cells.} \label{fig:Fig10}
\end{figure}

Figure \ref{fig:Fig10} presents microstrain versus dose for Voronoi and sheared cells, showing results from all simulations. Microstrain, reflecting lattice distortion, increases with peak broadening and can arise from grain boundaries, dislocations, voids, or interstitials. The microstrain was calculated by using Williamson-Hall on the MD data. The damage intervals in Figure \ref{fig:Fig10} match those in the experimental data. Note again the Voronoi-generated cell data stopping at 1 dpa due to the tetragonal distortion.

Figure \ref{fig:Fig10}(a) shows the microstrain for the five distinct simulations created through Voronoi tessellation with d$_{c}$ = 16.6 nm. The microstrain is obtained by computing diffraction patterns for the MD cells and performing a Williamson-Hall analysis. All simulations follow a very similar trend; there appears to be very little microstrain evolution between 0 and 0.001 dpa after which there is an increase in microstrain to 1 dpa. The d$_{c}$ = 5.2 nm cells in Figure \ref{fig:Fig10}(b) follow a similar trajectory, increasing after 0.001 dpa. However, it is noticeable that the microstrain in the d$_{c}$ = 5.2 nm cells begins to decrease after 0.1 dpa or 0.3 dpa for all cells, which is unlike the d$_{c}$ = 16.6 nm cells which see a continuous increase.

The sheared cell microstrain, shown in Figure \ref{fig:Fig10}(c), behaves very differently to the Voronoi cells. The starting microstrain is much larger for the sheared cells, being in the range of $\sim$ 0.012 to 0.017, compared to that of the Voronoi cells, which started out between $\sim$ 0.0015 and 0.0025. This may be explained by the large number of grains and existing material defects in the sheared cells at 0 dpa. Subsequently, the microstrain for the sheared cells decreases from 0 dpa to 5 dpa, which is unlike the behaviour shown for the Voronoi-created cells.

The MD results show the presence of an irradiation induced grain growth and defect reduction process for the nanocrystalline cells. Interestingly, a reduced swelling for nanocrystalline cells is only apparent when using the corresponding cells, at 0 dpa, as a reference, shown in Figure \ref{fig:Fig9}(a). Figure \ref{fig:Fig9}(b) suggests that all cells reach a similar amount of volumetric swelling irrespective of their initial starting microstructure for doses above 9 dpa, when referenced to a perfect single-crystalline cell. This suggests that the reported swelling resistance of NC materials is not a result of them being nanocrystalline. Rather, it is that starting material already contains a substantial amount of swelling, induced by the SPD process, so there is less scope for further swelling upon irradiation.

\subsection{Experimentally Observed Evolution of Irradiated Iron}

\begin{figure*}[p]
\centering
\includegraphics[width=0.9\linewidth]{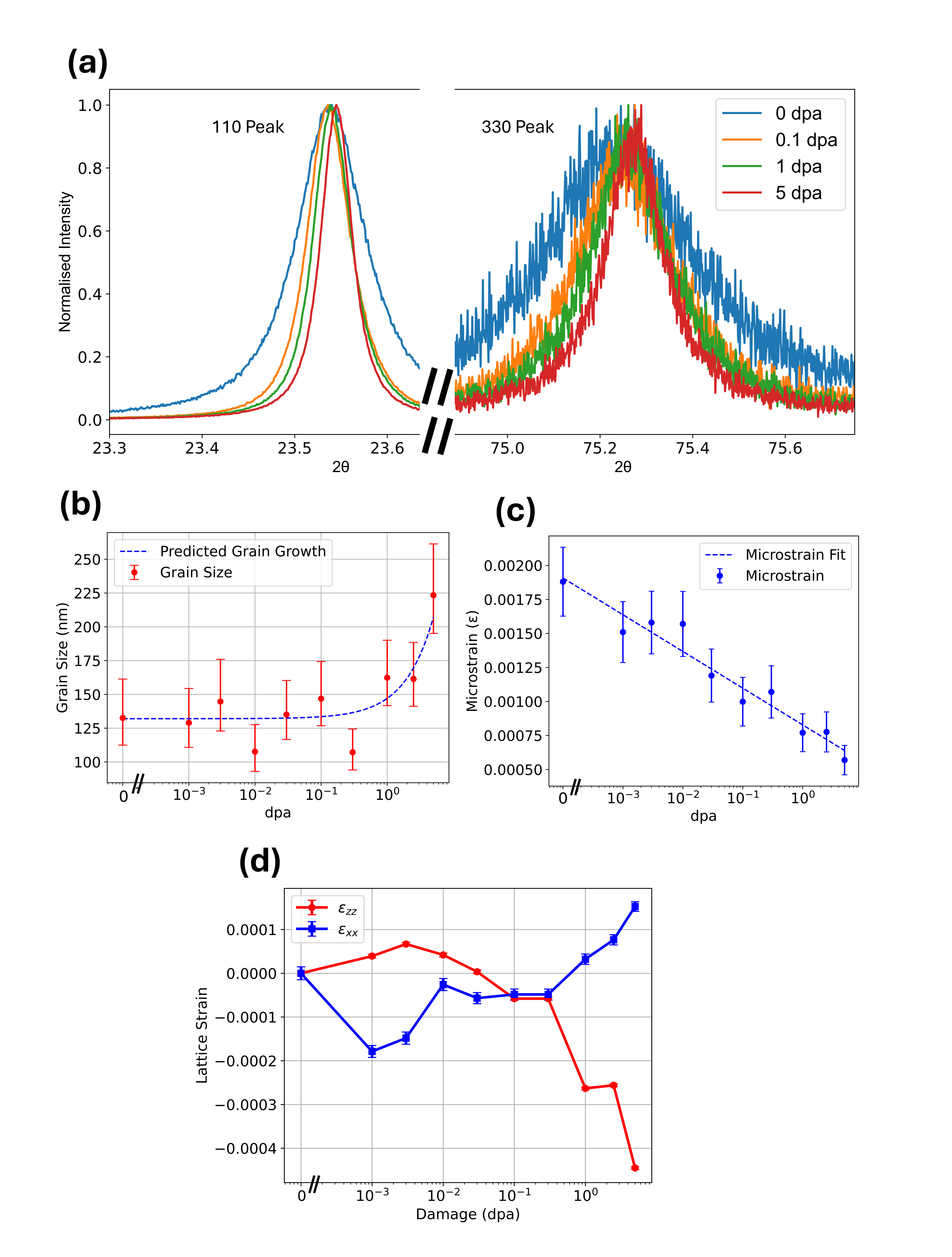}
\caption{Experimentally obtained data on ion-irradited Fe nanocrystalline samples. Line profiles obtained using grazing incidence X-ray diffraction; (a) Comparison of the [110] and [330] peaks for different damage levels, (b) Grain size as a functions of dose, (c) Microstrain as a function of dose, (d) In-plane ($\varepsilon_{xx}$) and out-of-plane ($\varepsilon_{zz}$) lattice strain as a function of dose.}
\label{fig:Fig11}
\end{figure*}

Experimental powder X-ray diffraction measurements were taken on ion irradiated, nanocrystalline iron. The results of the diffraction data analysis can be found in Figure \ref{fig:Fig11}. Figure \ref{fig:Fig11}(a) shows the [110] and the [330] peaks for the nanocrystalline iron as a function of dose, with the intensities being normalised. Peak narrowing with irradiation is evident, as well as a peak shift for both the low and high angle peaks. This suggests the presence of an irradiation induced annealing process as found in the MD simulations.

To further probe the XRD data, Williamson-Hall analysis was carried out \cite{WILLIAMSON195322}. The Williamson-Hall data can be directly converted to estimate grain size and microstrain within the material. The results of this are shown in Figures \ref{fig:Fig11}(b) and \ref{fig:Fig11}(c) which show the grain size and microstrain at different damage levels in the nanocrystalline iron, respectively.

Figure \ref{fig:Fig11}(b) shows that there appears to be little grain size evolution between 0 dpa, where the grain size is $\sim$ 130 nm, and 0.3 dpa, where the grain size is $\sim$ 110 nm. Whilst there is some variation in grain size up to 0.3 dpa, the error estimation shows that the grain sizes could be very similar for all samples. There appears to be a noticebale increase in grain size at 1 dpa and 2.5 dpa, being $\sim$ 165 nm, before there is a major increase in grain size to $\sim$ 220 nm at 5 dpa. A line of theoretical grain size increase is also included which is calculated using the volume exchange rate found in the MD simulations, shown in Figure \ref{fig:Fig6}(b) - further detail of this is given in the Discussion section. Fascinatingly, the experimental data correlates well to the theoretical grain size increase from MD simulations.

A reduction of microstrain with irradiation in the nanocrystalline iron samples is evident from Figure \ref{fig:Fig11}(c). At 0 dpa, the sample microstrain is $\sim$ 0.0018, and this reduces to $\sim$ 0.0005 at 5 dpa. This directly supports the notion that an irradiation induced annealing process is present for the ion-irradiated, nanocrystalline iron discs as this reduction in microstrain directly correlates to reduction in material defect density. Note the similarities In this plot compared to Figure \ref{fig:Fig10}(c), which shows microstrain reduction with dose for the sheared cell MD simulations.

Figure \ref{fig:Fig11}(d) shows the in-plane lattice strain, denoted by $\varepsilon_{xx}$, and the out-of-plane lattice strain, denoted by $\varepsilon_{zz}$, with irradiation, for the nanocrystalline Fe samples. The reference for these measurements is the 0 dpa sample, that is the data shows the change in lattice strain as a function of irradiation. Overall, the strains are very small with the in-plane lattice strain reducing to $\sim$ -0.0002 at 0.001 dpa, before increasing and returning close to 0 as shown in Figure \ref{fig:Fig10}(d). Similar behaviour has been previously observed for ion-irradiated Tungsten \cite{DAS20181226, HOFMANN2015352} which showed in-plane strain to be small due to the constraint imposed by the unirradiated material beyond the implanted layer. Interestingly, after 0.3 dpa, the in-plane lattice strain begins to increase before reaching $\sim$ 0.00015 at 5 dpa. It is not immediately obvious whether the in-plane lattice strain will continue to evolve with irradiation or will continue to fluctuate around 0. 

Figure \ref{fig:Fig11}(d) also clearly shows the reduction of out-of-plane lattice strain with irradiation. Whilst initially, there is an increase in out-of-plane lattice strain to 0.00005 strain at 0.003 dpa, after this point, the strain reduces and ultimately reaches -0.00045 at 5 dpa. It is interesting that $\varepsilon_{xx}$ and $\varepsilon_{zz}$ almost follow opposite evolutions. Whilst $\varepsilon_{xx}$ initially decreases, $\varepsilon_{zz}$ initially increases. At 0.1 dpa and 0.3 dpa, both $\varepsilon_{xx}$ and $\varepsilon_{yy}$ are almost identical, with a net reduction in strain ($<$ 0). At 0.3 dpa, $\varepsilon_{xx}$ increases whilst $\varepsilon_{zz}$ decreases, with the overall strain being negative in the material, as the absolute values for $\varepsilon_{zz}$ are higher than that for $\varepsilon_{xx}$.

This behaviour is opposite to that which is observed for ion-irradiated, coarse grained Fe as shown by Song \textit{et al.} \cite{SONG2024154998}. Song utilises the same EFDA material as in this study, keeping them coarse grained with a grain size of $\sim 187 \pm 150 \ \mu m$, and irradiating to various doses with 20 MeV $Fe^{4+}$ ions at the Helsinki Accelerator Laboratory. With these parameters, Song \textit{et al.} found that the out-of-plane lattice strain increased as a function of dose to $\sim 2 \times 10^{-4}$ at 6 dpa. Clearly, the behaviour observed for the coarse grained and nanocrystalline cells is very different. 

For BCC Fe, vacancies have a negative relaxation volume ($\Omega_{vac} = -0.220$), whilst self-interstitials have a positive relaxation volume ($\Omega_{int} \sim 1.6-1.8$) \cite{LeoMaVolume}, which means that the Frenkel pair has a positive relaxation volume. As such, the reduction in overall lattice strain with irradiation in Figure \ref{fig:Fig11}(d) suggests that there is a removal of interstitial defects within the system or a greater detention of vacancies than interstitials. 

\section{Discussion}

\begin{figure*}[t]
\includegraphics[width=\linewidth]{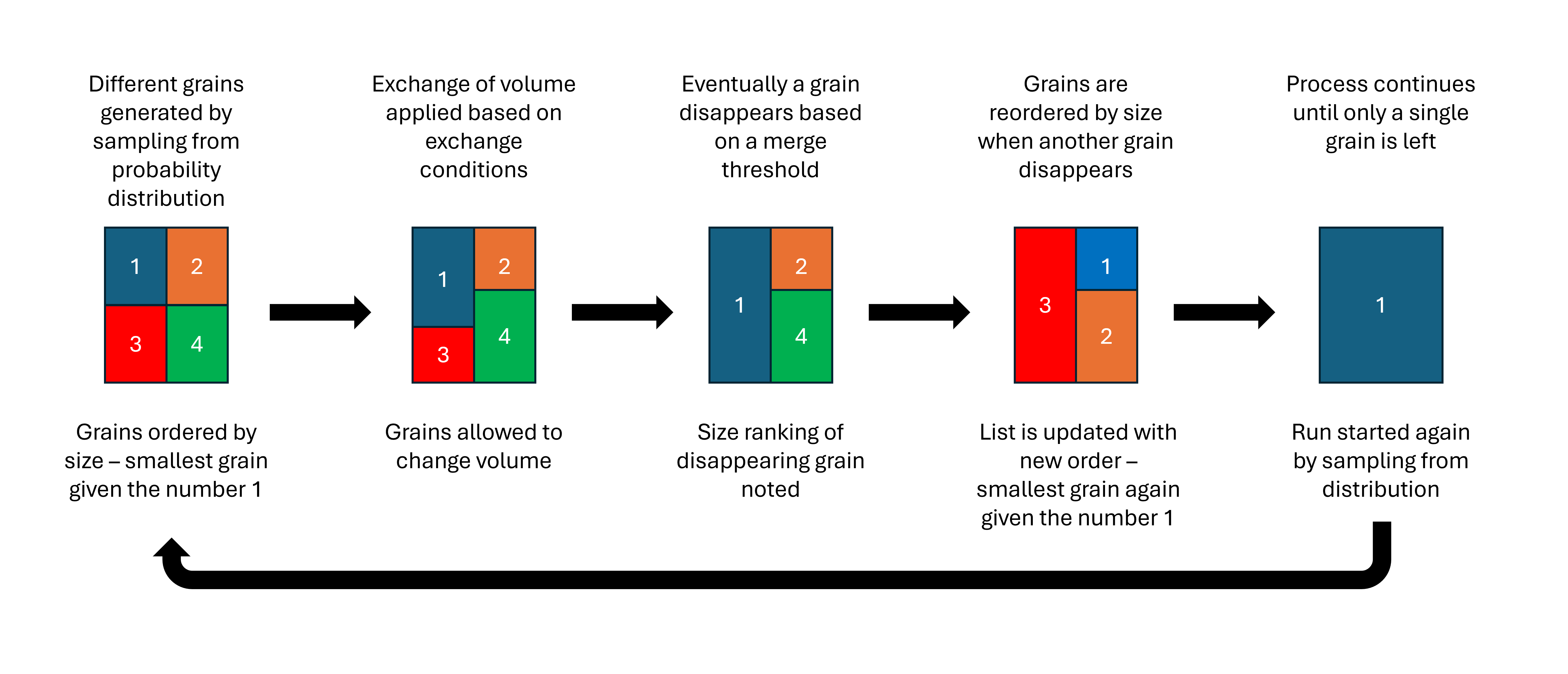}
\caption{Representation of the Toy Model}
\label{fig:Fig12}
\end{figure*}

The molecular dynamics results reveal the presence of a grain coarsening process when irradiating nanocrystalline iron, as shown in Figure \ref{fig:Fig6}. The experimental data on ion-irradiated nanocrystalline iron shows a similar trend, as seen in Figure \ref{fig:Fig11}(b). The MD results are used to predict the grain growth seen in the experimental data in the following way: Williamson-Hall analysis is firstly used to calculate the radius of grains in the sheared cells. The data between 0.03 dpa and 1 dpa is used to fit r as a function of dpa, shown in Figure \ref{fig:Fig6}(b), with the function being

\begin{equation} \label{eq3}
    r = 7.52 \cdot dpa + r_{0}
\end{equation}
\\

Here, $r_{0}$ = 3.9715 nm, and represents the mean starting radius of the sheared cells at 0.03 dpa. Below 0.03 dpa grain size distribution varies substantially between simulations, and there is also increased uncertainty in the grain size determined from WH analysis, as seen in Figure \ref{fig:Fig6}(c). Above 1 dpa, the simulation cell size starts to become important. Indeed once the simulation cell has become single crystalline, no further grain growth is possible, as seen in Figure \ref{fig:Fig6}(b).

Note, that the relationship between radius and dose appears to be linear in this case, and in fact the $R^{2}$ value of fitting the linear function is 0.991, giving good confidence that Equation (\ref{eq3}) is a good estimate of the MD cell behaviour. However, there is no reason to suggest that this behaviour should be linear, and another model is proposed in Appendix E which clarifies why it appears that the radius increases linearly with dose.

Equation (\ref{eq3}) was scaled to the experimental data by setting $r_{0}$ to 132 nm, and the grain growth prediction, derived from MD data, is overlaid on the experimental data in Figure \ref{fig:Fig11}(b). The anticipated grain growth correlates well with the experimental data, giving us confidence in the physics of the MD simulations, which have seemingly predicted the grain growth of real nanocrystalline, self-ion irradiated iron.

The microstrain in the experimental samples decreases with irradiation, as shown in Figure \ref{fig:Fig11}(c) and a very similar behaviour is observed in the irradiation simulations of the sheared cells, in Figure \ref{fig:Fig10}(c). This gives confidence to the microstructural evolution mechanisms found in the simulation data as being representative of experimental data. The reduction in microstrain relates to a reduced distortion in the material lattice, which is attributed to an irradiation induced annealing process. Note that the microstrain evolution for the Voronoi cells, shown in Figures \ref{fig:Fig12}(a) and \ref{fig:Fig12}(b), is rather different to the experimental data suggesting that nanocrystalline MD cells created through shear are more representative of real material.

The simulation data also shows that the dislocation density is lower at high doses for initially nanocrystalline cells when compared to initially pristine cells, notwithstanding the fact that all initially nanocrystalline cells become single crystalline beyond $\sim$ 2 dpa (Figure \ref{fig:Fig7}). This was attributed to the presence of large dislocation networks for the pristine cells compared to the presence of small loops for the initially nanocrystalline cells as confirmed by the histograms in Figure \ref{fig:Fig8}. Since the pristine cells have no grain boundaries to constrain dislocation formation and growth, large networks can form which are then too energetically costly to remove. Conversely, the presence of GBs in the initially NC cells constrained the formation and growth of dislocations, which resulted in the presence of small dislocation loops even as the NC cells become single crystalline at high damage levels. 

The sheared cells begin with a large dislocation density which rapidly reduces with irradiation until it is less than the pristine case, as shown in Figure \ref{fig:Fig7}(a). The reduction in microstrain found in the experimental data (Figure \ref{fig:Fig12}(c)) also suggests a reduction of defect density within the material. It is difficult to ascertain the nature of the defects which are being removed experimentally, but it is encouraging that the reduction of defects with irradiation can be seen in both simulations and experiments.

A resistance to swelling is also observed in the simulated data, presented in Figure \ref{fig:Fig9}(a). Whilst the pristine samples continually swell with irradiation, and appear to be increasing with dose even at 5 dpa, the nanocrystalline samples show a reduction in swelling after an initial increase, before plateauing. In fact, the sheared cells exhibit a reduction in swelling below that of the original sheared configurations at 0 dpa. This swelling resistance has been previously shown in CRA simulations on nanocrystalline Tungsten \cite{LeoMaPRM}.  

Swelling was also calculated when referenced to a perfect single crystalline cell for all nanocrystalline configurations (Figure \ref{fig:Fig9}(b)). This generated an interesting result in that, at doses above 1.5 dpa, all the nanocrystalline cells converged to a similar swelling value irrespective of the initial configuration. This suggests that at high damage levels the material forgets its initial configuration and converges at similar stable single crystalline states. A single instance of all cell configurations was run until 10 dpa to determine whether the swelling of the pristine cell will reach that of the initially nanocrystalline cells. Figure \ref{fig:Fig9}(c) shows that after 9 dpa, the swelling of the pristine cell plateaus at a similar value as the nanocrystalline cells, suggesting a late-blooming phase in the material's evolution. This further suggests that, from a swelling perspective, and at high doses above 9 dpa, the material forgets its initial configuration and converges to a critical drive state irrespective of initial material microstructure.

It is apparent, from both the simulated and experimental data, that an irradiation-driven grain growth process is present. This raises the following questions: \textbf{What is special about the grain that grows? What characteristics drive growth?}

\begin{table*}[t]
\centering
\begin{tabular}{ |p{3cm}||p{3cm}|p{3cm}|p{3cm}|  }
 \hline
 \multicolumn{4}{|c|}{$\chi^{2}$ Testing} \\
 \hline
 Name of Exchange Mechanism&Sheared Cells &d$_{c}$ = 16.6 nm cells&d$_{c}$ = 5.2 nm cells\\
 \hline
 RAN   & 109.348 &2.432&  21.535\\
 RAV& 54677.138 & 7593.479 &338836.067\\
 RAB& 4682.300 & 406853.721 & 148398.524\\
 ABV& 724.941 & 42612.018 & 66010.496\\
 EDR& 372852.560 & 5677.142 & 12403.484\\
 EDA& 6562.140 & 1200242.769 & 44195.629\\
 Critical $\chi^{2}$ Value& 113.145 & 9.488 & 30.144\\
 \hline
\end{tabular}
\caption{$\chi^{2}$ testing of different proposed grain growth mechanisms.}
\label{table:Table1}
\end{table*}

To answer this question, a \textit{Toy Model} was created, which is shown illustratively in Figure \ref{fig:Fig12}. The steps to create this model are outlined as such:

\begin{enumerate}
    \item The input cell volume is taken as 11,852,352 \AA$^{3}$, which reflects the cell volume of the MD simulation cells.
    \item This volume is distributed into grains by sampling from Equation \ref{eq1} and Equation \ref{eq2}, and utilising the Newton-Raphson method e.g. if we want to look at a 5 grain Voronoi cell, we break the total volume down into 5 equal pieces by sampling from Equation \ref{eq1}.
    \item The grains are ranked in order of size, with 1 being the smallest, and the total number of grains being the largest e.g. for the 5 grain Voronoi case, 1 will be the smallest and 5 will be the largest.
    \item These grains are allowed to exchange volume based on different exchange mechanisms, listed below, which correspond to different hypotheses of grain growth characteristics. This is shown illustratively in Figure \ref{fig:Fig12}.
    \item Once a grain drops below a volume threshold, it is removed from the system and its original rank is noted e.g. if the second smallest grain disappears, then we tally the number 2.
    \item The grains are again ranked from smallest to largest and the process is repeated until a single grain remains.
    \item This is repeated for each exchange hypothesis for 1 million runs. \item The data is then converted into a probability that the n-th largest grain will be the next one to disappear based on a particular exchange mechanism.
    \item Next, the disappearance order of grains in the MD simulation data is also counted, whereby all the grains for all distinct nanocrystalline simulations are ranked in terms of volume and disappearing grains are noted based on the rank of their volume.
    \item The MD and Toy Model datasets are directly compared using $\chi^{2}$ testing, to determine whether any of the hypotheses from the Toy Model can be true for the MD data.
\end{enumerate}

A convergence study was carried out to balance computational efficiency and to ensure that the smallest possible volume was exchanged between steps, the results of which are provided in Appendix F.

The exchange mechanisms that are employed in the Toy Model are based on certain hypotheses. In each iteration step, the model chooses two grains, A and B, with one grain receiving the volume and the other one losing it based on the exchange mechanisms. The first hypothesis is that the grain growth is completely stochastic whereby the energies and dose rates of the cascades are so high that the grains will grow at random. The second hypothesis is that out of the two grains, A and B, the largest grain will always grow. This is the physical representation of long-range diffusion whereby larger grains attract atoms from larger distances. Lastly, the third hypothesis is that, out of the selected grains A and B, the one with the lowest elastic energy density always grows. The elastic energy density of the grains in the Toy Model is approximated by:

\begin{equation} \label{eq5}
    U \propto \frac{1}{d} = V_{Grain}^{-1/3}
\end{equation}
\\

This hypothesis proposes that the grain growth is driven by energy minimisation. The full list of exchange mechanisms are listed below:

\begin{enumerate}
    \item Grain growth is truly random, scaled from a normal distribution - RAN
    \item Grain growth is random but is scaled by the volume of the grain chosen to increase in size - RAV
    \item Grain growth is random but grain A is always bigger than grain B, and grain A always grows - RAB
    \item Grain A is always bigger than grain B and grain A always grows with the growth being proportional to the volume of A - ABV
    \item Grain growth is random but is always scaled by the inverse of the elastic energy density of the grain that grows - EDR
    \item Grain A always has lower elastic energy density than grain B, and the growth of grain A is proportional to the inverse of its elastic energy density - EDA
\end{enumerate}

A direct comparison is made between the Toy Model and the MD data through $\chi^{2}$ testing. A significance level of 0.05 is taken to calculate the critical $\chi^{2}$ values. The degree of freedom is the number of grains for each case minus 1. Note that due to the low availability of data in the d$_{c}$ = 16.6 nm MD simulations (only 5 grains per cell), the Yates correction \cite{YatesCorrection} is applied. The results are shown in Table \ref{table:Table1}.

The critical values for the $\chi^{2}$ test are found in the bottom row. If a mechanism shows a value below this, then it is deemed to come from the same distribution and the null hypothesis is accepted. Similarly, if the $\chi^{2}$ value is above the critical value, then the null hypothesis is rejected and the distributions are not deemed to be the same. Clearly, as Table \ref{table:Table1} shows, the only hypothesis that can be accepted is RAN, which states that the grain growth is completely random and is scaled by sampling from a normal distribution. This is interesting as it suggests that the initial volume and elastic energy density of grains do not play a significant role in determining which grains grow and which shrink.  An interesting remark in the context of textured materials was made by Novikov \cite{NOVIKOV19991935} who pointed out that simply having more grains of a certain orientation will ensure that these grains are the ones that end up growing.

\section{Conclusion}

Using MD simulations, we study the evolution of nanocrystalline iron when subjected to cascade damage up to 5 dpa and compare this to initially pristine iron cells. Nanocrystalline cells created through Voronoi tessellation, are found to have no texture. On the other hand, nanocrystalline cells created by  plastic shearing deformation, are strongly textured. Experimentally, we carry out a Williamson-Hall analysis on line profiles obtained from XRD measurements on nanocrystalline, self-ion irradiated iron discs. As such, a combined atomistic modelling and experimental study on irradiated nanocrystalline iron is possible.

Our results allow the following conclusions to be reached:

\begin{itemize}
  \item An irradiation induced grain growth process is seen for the initially nanocrystalline cells, irrespective of whether they were created through Voronoi tessellation or through plastic shearing deformation. After $\sim$ 2 dpa, all initially nanocrystalline cells became single crystalline. This growth in grain size is experimentally validated through Williamson-Hall analysis. A constant rate of grain radius increase with dose was determined from MD simulations. This correlated well with experiments and could predict the experimental grain growth process.
  \item A Toy Model is created, based of numerous grain growth hypotheses, to determine whether there were any special characteristics associated with the grains that grow in our MD simulations. Based on this, we conclude that the grain growth is consistent with random growth. This suggests that the initial volume and elastic energy density of grains have little influence on determining which grains grow and which ones shrink.
  \item An irradiation induced annealing process in the MD simulations is also shown for the sheared cells through a microstrain analysis. This is further validated by a microstrain analysis of the experimental data. Interestingly, the cells created through Voronoi tessellation do not experience the same trends in microstrain reduction and are not representative of the experimental data.
  \item Initially nanocrystalline cells created through Voronoi tessellation show a lower dislocation density increase with dose than initially pristine cells above $\sim$ 0.02 dpa, and remain lower until 5 dpa, plateauing at below half the value of the initially pristine cells. The sheared cells start with a large dislocation density, due to the nature of their creation, before reducing and plateauing at a similar level to the Voronoi-created cells. A line length analysis comparing the sheared and initially pristine cells shows that the grain boundaries in the sheared cells constrain the formation of large dislocation networks as present in the initially pristine cells, which become too energetically difficult to remove.
  \item An irradiation induced swelling resistance is observed in the initially nanocrystalline cells when referenced against the same cells at 0 dpa. However, when referenced against an initially pristine iron cell, it is found that, irrespective of the initial cell configuration, all initially nanocrystalline cells plateaued at similar swelling values. In fact, a single instance of each cell was simulated to 10 dpa and this showed that even the initially pristine cells plateaued at a similar swelling value to the initially nanocrystalline cells beyond $\sim$ 9 dpa. This indicates that the reported swelling resistance of NC materials is not inherently due to their nanocrystalline nature. Instead, it stems from the fact that the starting material already exhibits significant swelling caused by the SPD process, leaving limited potential for additional swelling under irradiation.
\end{itemize}

\section{Data Availability}

All input scripts and simulation data presented in the current work are available at \textit{A link will be provided after the review process and before publication.} 

\section*{Acknowledgements}

The authors gratefully acknowledge funding from the Department of Engineering Science at the University of Oxford. This work has been carried out within the framework of the EUROfusion Consortium, funded by the European Union via the Euratom Research and Training Programme (Grant Agreement No 101052200 — EUROfusion) and by the EPSRC (Grant number EP/W006839/1). To obtain further information on the data and models underlying this paper please contact PublicationsManager@ukaea.uk. Views and opinions expressed are those of the author(s) only and do not necessarily reflect those of the European Union or the European Commission. Neither the European Union nor the European Commission can be held responsible for them. The authors acknowledge the use of the ARCHER2 UK National Supercomputing Service (https://www.archer2.ac.uk) under project e804 and associated support services provided by the ARCHER2 Service Desk in the completion of this work. This work also used the Cambridge Service for Data Driven Discovery (CSD3) (www.csd3.cam.ac.uk).

\appendix
\section{SRIM Recoil Energies}

\begin{figure}
\begin{subfigure}{0.5\textwidth}
\includegraphics[width=\linewidth]{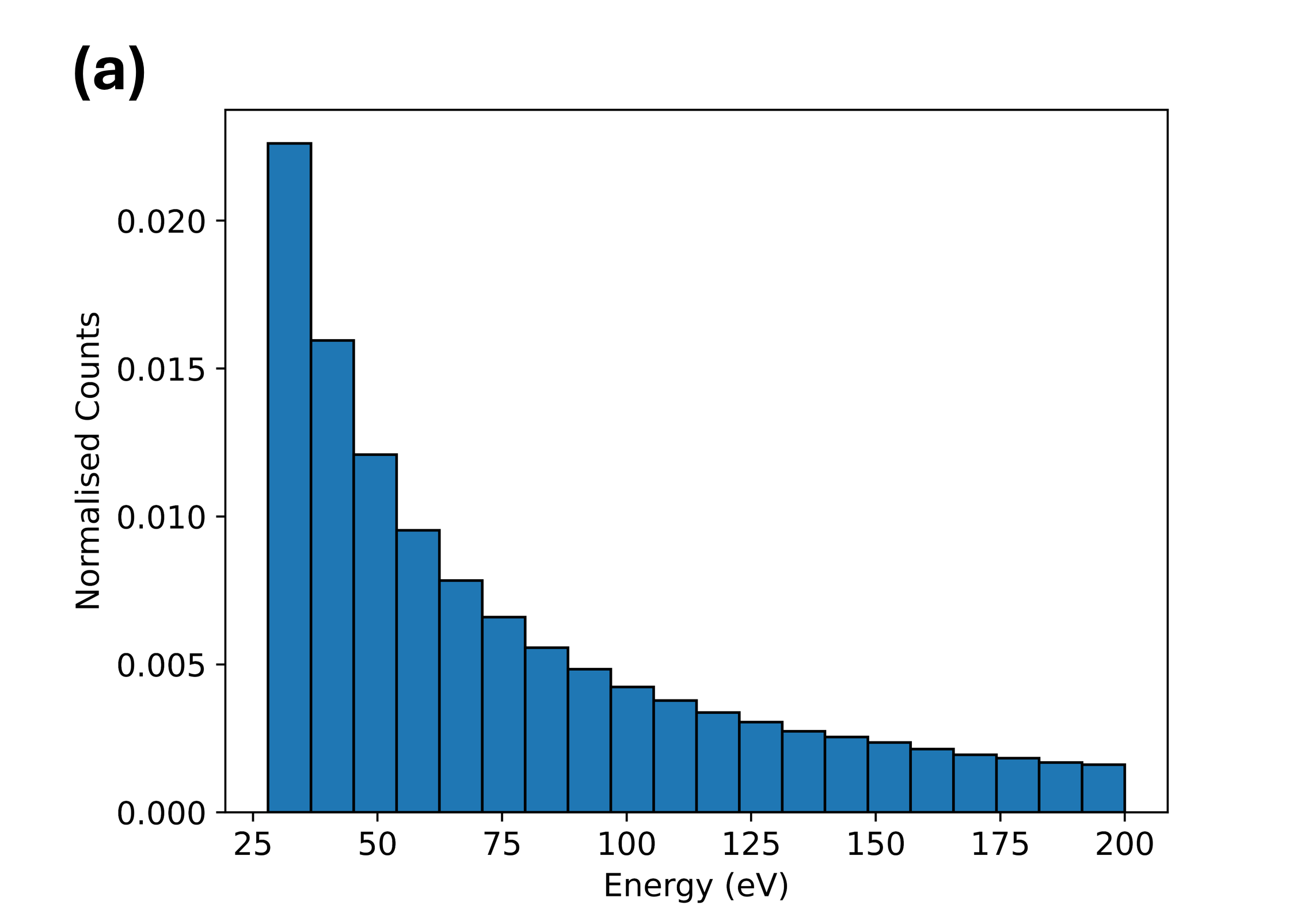}
\end{subfigure}
\begin{subfigure}{0.5\textwidth}
\includegraphics[width=\linewidth]{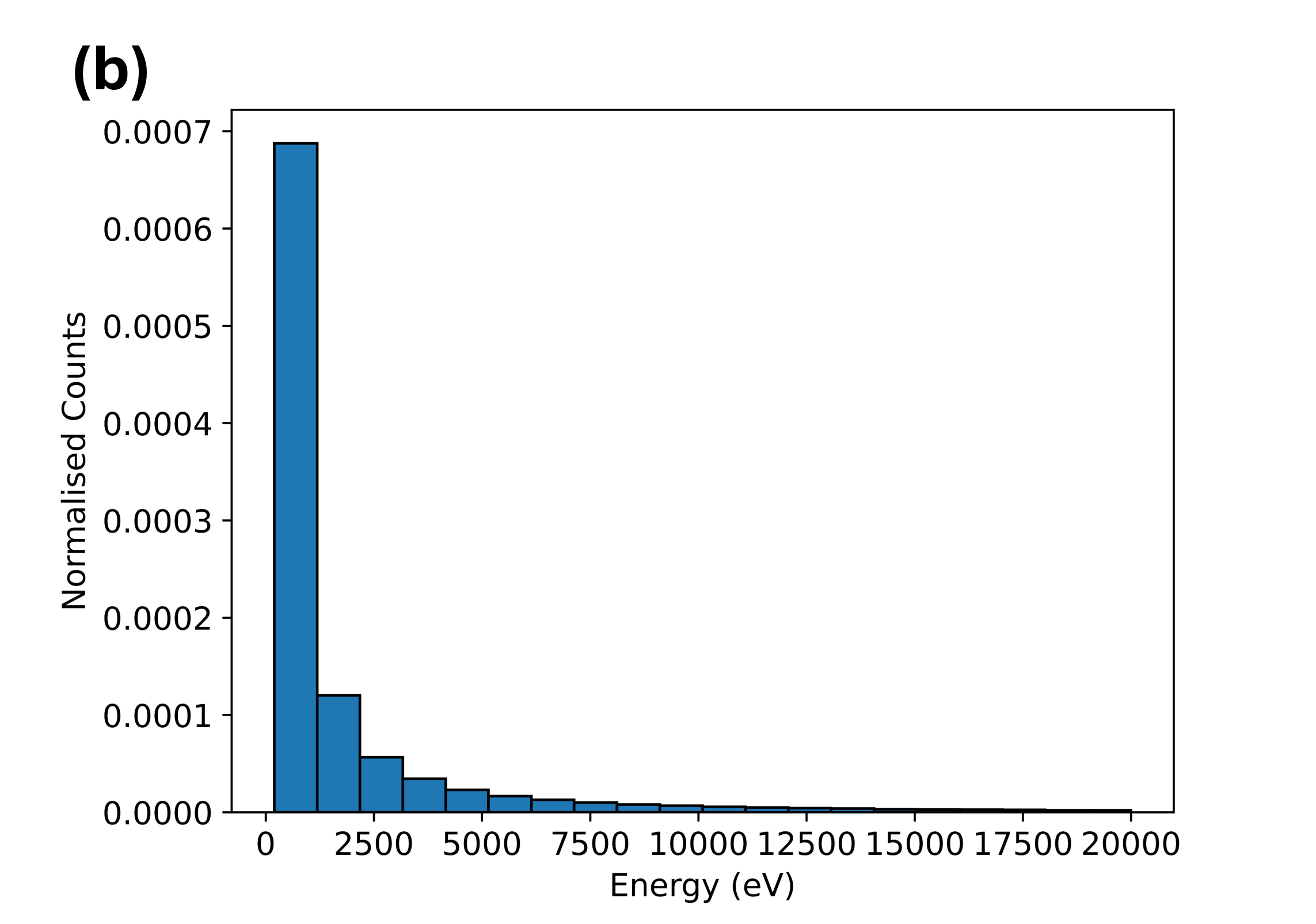}
\end{subfigure}
\caption{Normalised histograms of PKA energies from SRIM; (a) energies between 30 - 200eV, (b) energies between 200 eV - 20 keV.} \label{fig:Fig13}
\end{figure}

\begin{figure*}[t]
\renewcommand{\thefigure}{B.14}
\begin{subfigure}{1.0\textwidth}
\includegraphics[width=\linewidth]{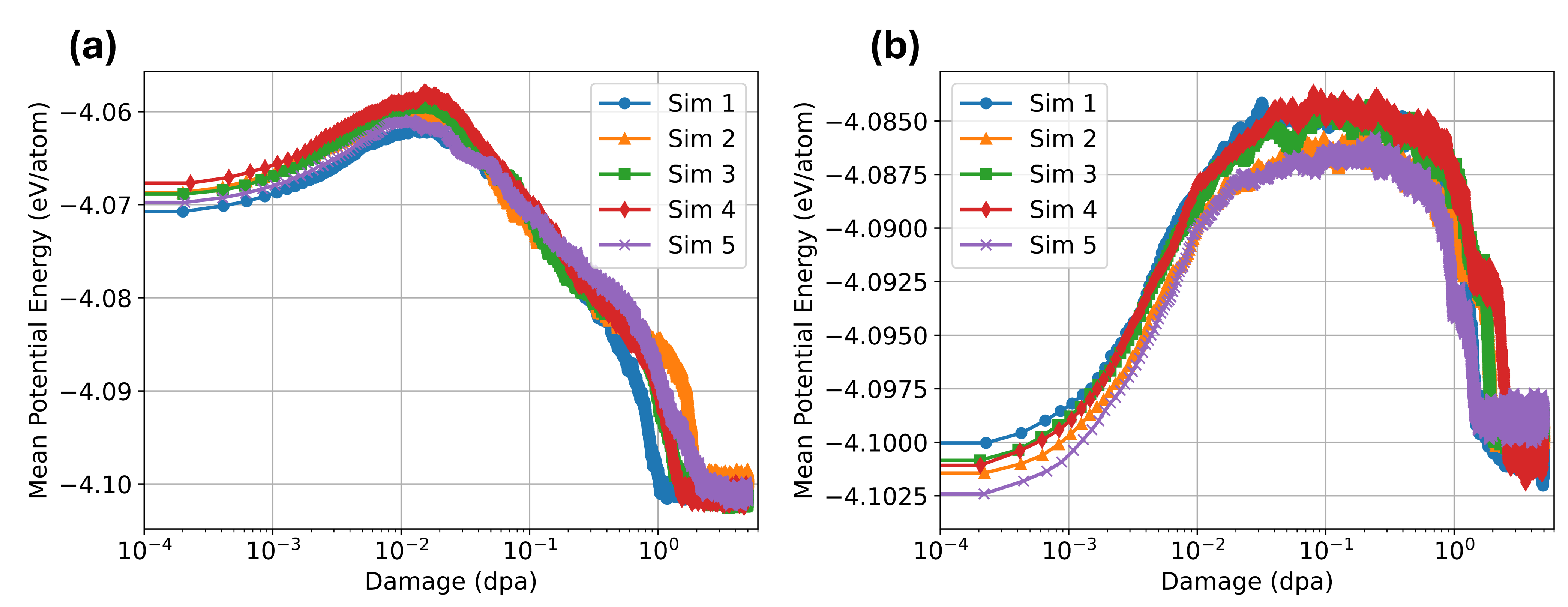}
\end{subfigure}\
\begin{subfigure}{1.0\textwidth}
\includegraphics[width=\linewidth]{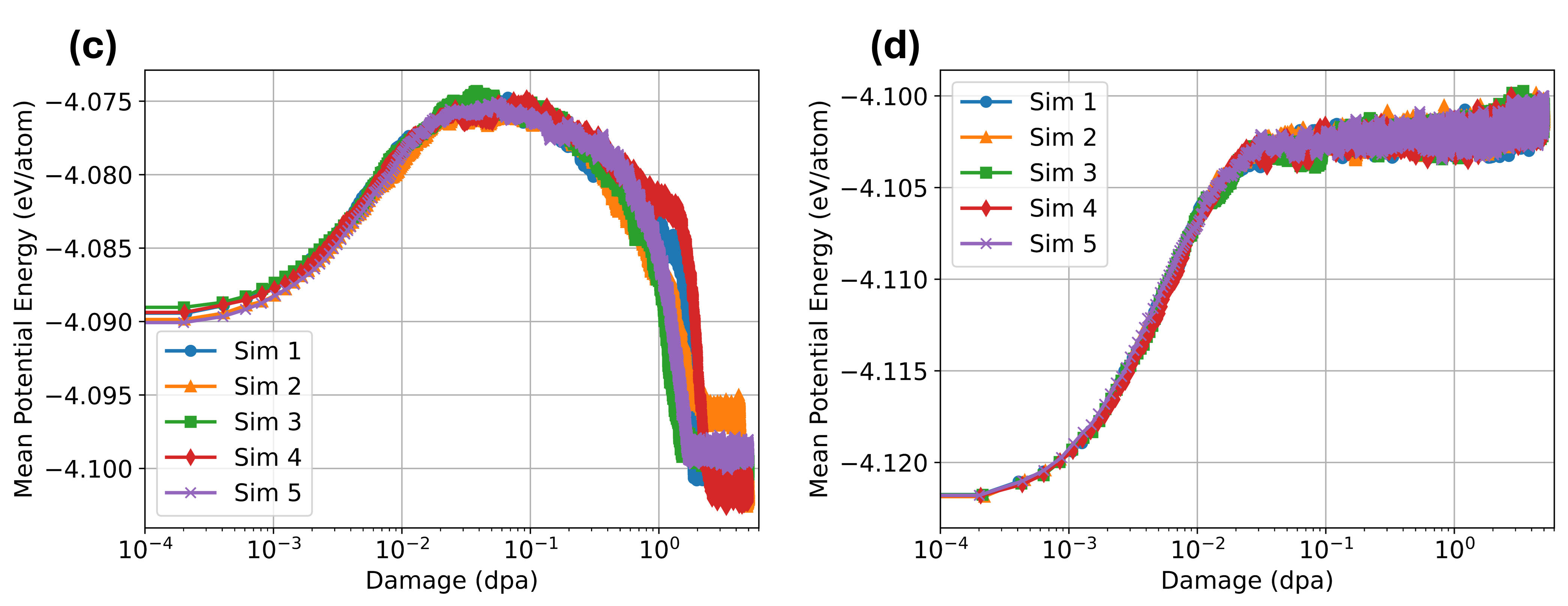}
\end{subfigure}\
\caption{Potential energies as a function of dose for five distinct simulations for each initial material condition; (a) \textit{sheared}, (b) d$_{c}$ = 16.6 nm, (c) d$_{c}$ = 5.2 nm, (d) pristine cells.}
\label{fig:Fig14}
\end{figure*}

The histogram of recoil energies produced by SRIM, as discussed in the Methods section, can be found in Figure \ref{fig:Fig13}. Figure \ref{fig:Fig13}(a) depicts the histogram for the low energies, between 30 and 200 eV, whilst Figure \ref{fig:Fig13}(b) depicts the high energy histogram, between 200 eV and 20 keV. The range from which PKA energies were drawn is 50 eV - 20 keV.

It can be seen that the majority of the PKA energies are low, with only a few high energy recoils obtained from SRIM. As such, it is reasonable to assume that most cascade energies are between 50 eV and 200 eV. It is possible to consider higher energy cascades within the simulations, for which the damage would be calculated accordingly and the same exclusion radius rules would apply.

\section{Simulation Repeatability}

The mean potential energy (eV/atom) for the five distinct simulations is shown for each cell type in Figure \ref{fig:Fig14}. The results show that the mean potential energies of the five distinct simulations follow near identical trajectories with irradiation. This points to the repeatability of these molecular dynamics simulations.

The data in Figure \ref{fig:Fig6} also points to the repeatability of these simulations. In Figure \ref{fig:Fig6}, the number of grains as a function of dose is shown for the nanocrystalline cells. Whilst the standard deviation of the sheared cells begins large, with increased dose, the grain number becomes almost homogeneous over the five distinct simulations for each case, as shown by the standard deviation. This suggests that the grains evolve at similar rates throughout all simulations.

\section{Instrument Broadening}

\begin{figure}
\includegraphics[width=\linewidth]{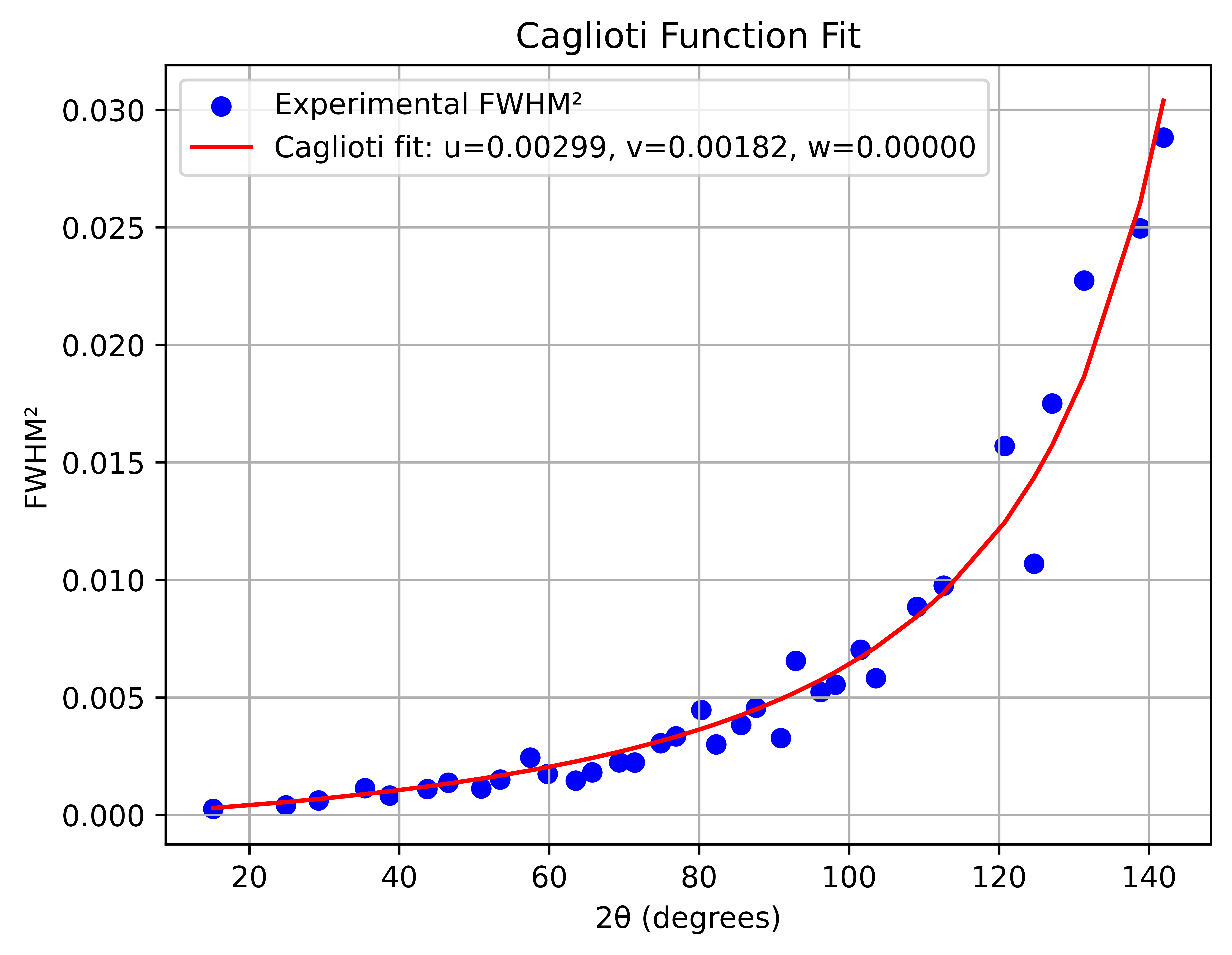}
\caption{Caglioti function fitted to Silicon 640c calibration sample to account for instrument broadening.} \label{fig:Fig15}
\end{figure}

The Diamond Light Source I11 MAC detectors were used to collect the presented diffraction data. These are high resolution detectors , designed to minimise instrumental broadening. A Silicon 640C sample was measured for calibration at multiple times during the experiment. The silicon data was subsequently used to fit an instrument broadening function based on the work of Caglioti \textit{et al.} \cite{CAGLIOTI1958223}. This fitting is shown in Figure \ref{fig:Fig15} and results in the following Caglioti function

\begin{equation} \label{eqC1}
    FWHM^{2} = 0.00299\tan^{2}(\theta) + 0.00182\tan(\theta) + 0.0000
\end{equation}

This is the function which corresponds to the instrumental broadening and is subtracted from the Williamson-Hall fitting to remove the instrumental peak-width contribution.

\section{Cell Size Limitation}

\begin{figure*}
\includegraphics[width=\linewidth]{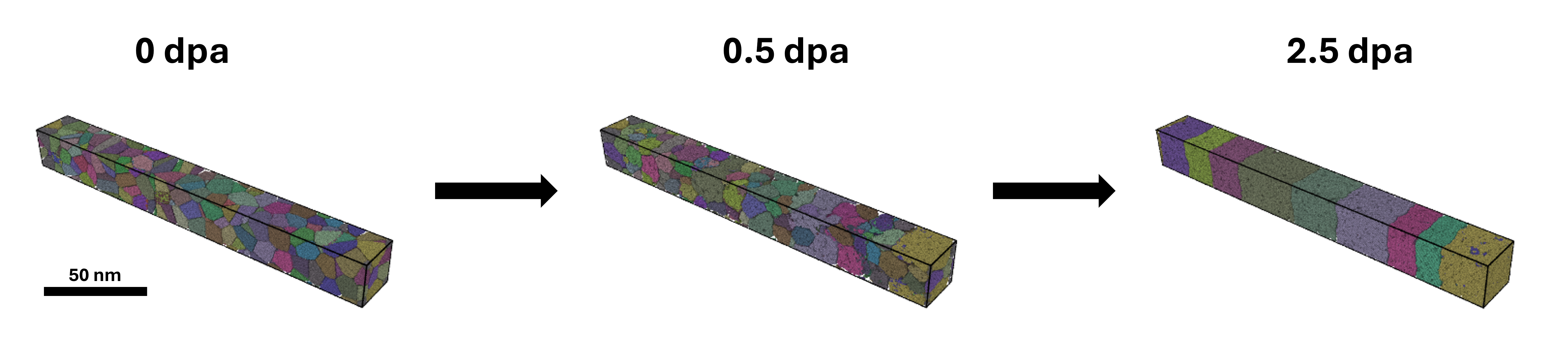}
\caption{Grain coarsening visualisation of needle simulation}
\label{fig:Fig16}
\end{figure*}

As shown in the results section, all nanocrystalline cells became single crystalline under irradiation. As such, it is difficult to determine whether there is any size limit for the grain coarsening process or whether the coarsening will continue with increasing dose. To better understand this question, a simulation was carried out on a needle shaped $\sim$ 5 million atom cell, with an aspect ratio of around 1:9. This cell was created through Voronoi tessellation and initially had 100 grains. It was irradiated through cascade simulations to 2.5 dpa, which is roughly the dose at which all nanocrystalline 1 million atom cells had become single crystalline.

\begin{figure}
\includegraphics[width=\linewidth]{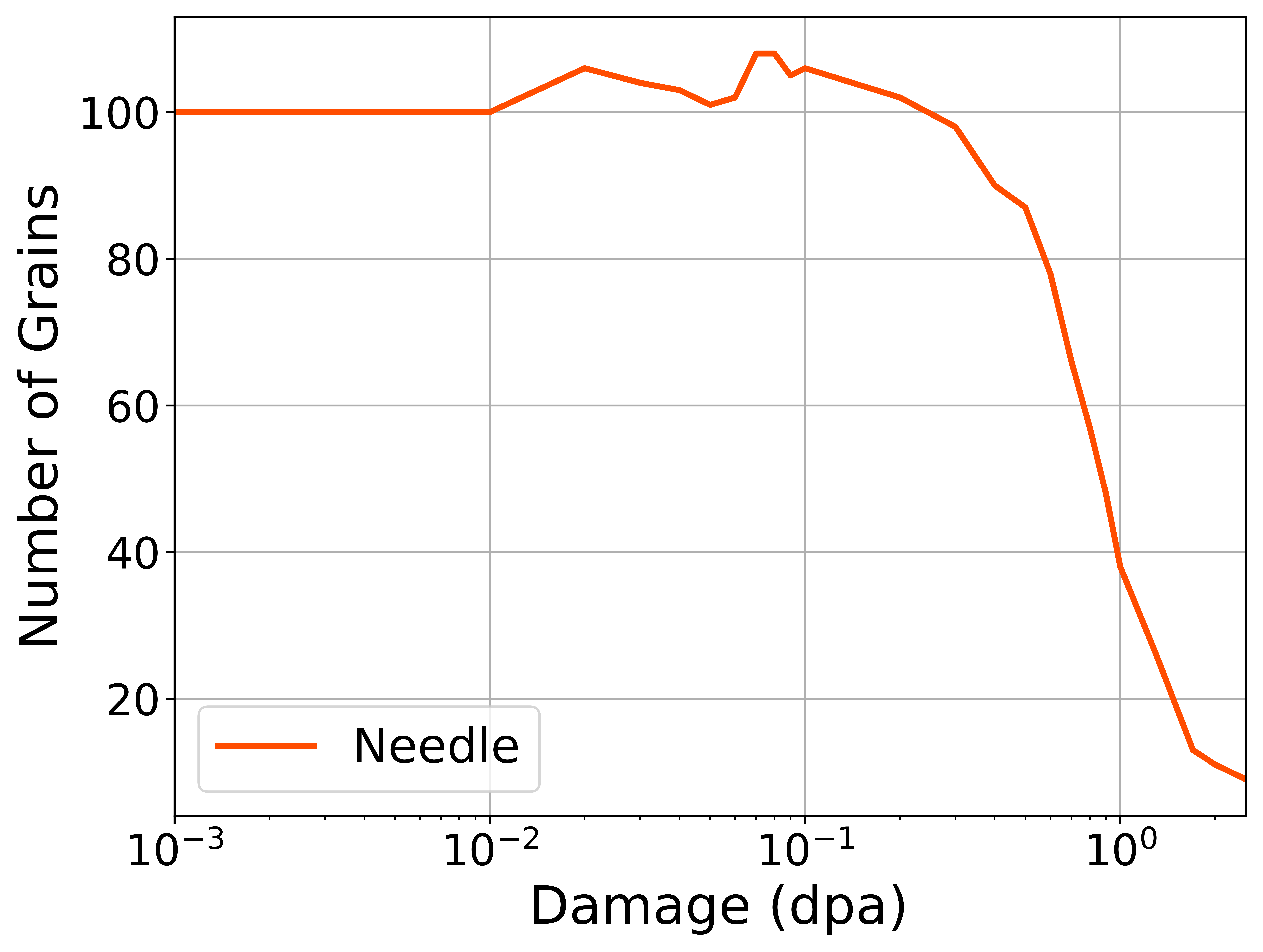}
\caption{Number of grains with dose for needle simulation}
\label{fig:Fig17}
\end{figure}

Figure \ref{fig:Fig16} depicts the grain coarsening process for the nanocrystalline needle. Interestingly, at 2.5 dpa, there are still many grains present which take a columnar form and are around 25 - 50 nm in size. This is larger than any dimension in the 1 million atom cells and may be the reason why this upper limit is not captured. The grain growth is also shown in Figure \ref{fig:Fig17} which shows number of grains as a function of dose for the needle simulation. The number of grains remains at 100 up to 0.01 dpa after which it increases, peaking at $\sim$ 110 grains at 0.07 dpa, before reducing to 9 grains at 2.5 dpa. It is clear from Figures \ref{fig:Fig16} and \ref{fig:Fig17} that the nanocrystalline needle does not become single crystalline at 2.5 dpa, contrary to the 1 million atom nanocrystalline cells. In the future, it would be interesting to carry out cascade simulations with larger cell size to try to capture whether there is a limiting grain size of irradiation however, this is not considered here.

\section{Radius Increase Computational Model}

\begin{figure}
\begin{subfigure}{0.5\textwidth}
\includegraphics[width=\linewidth]{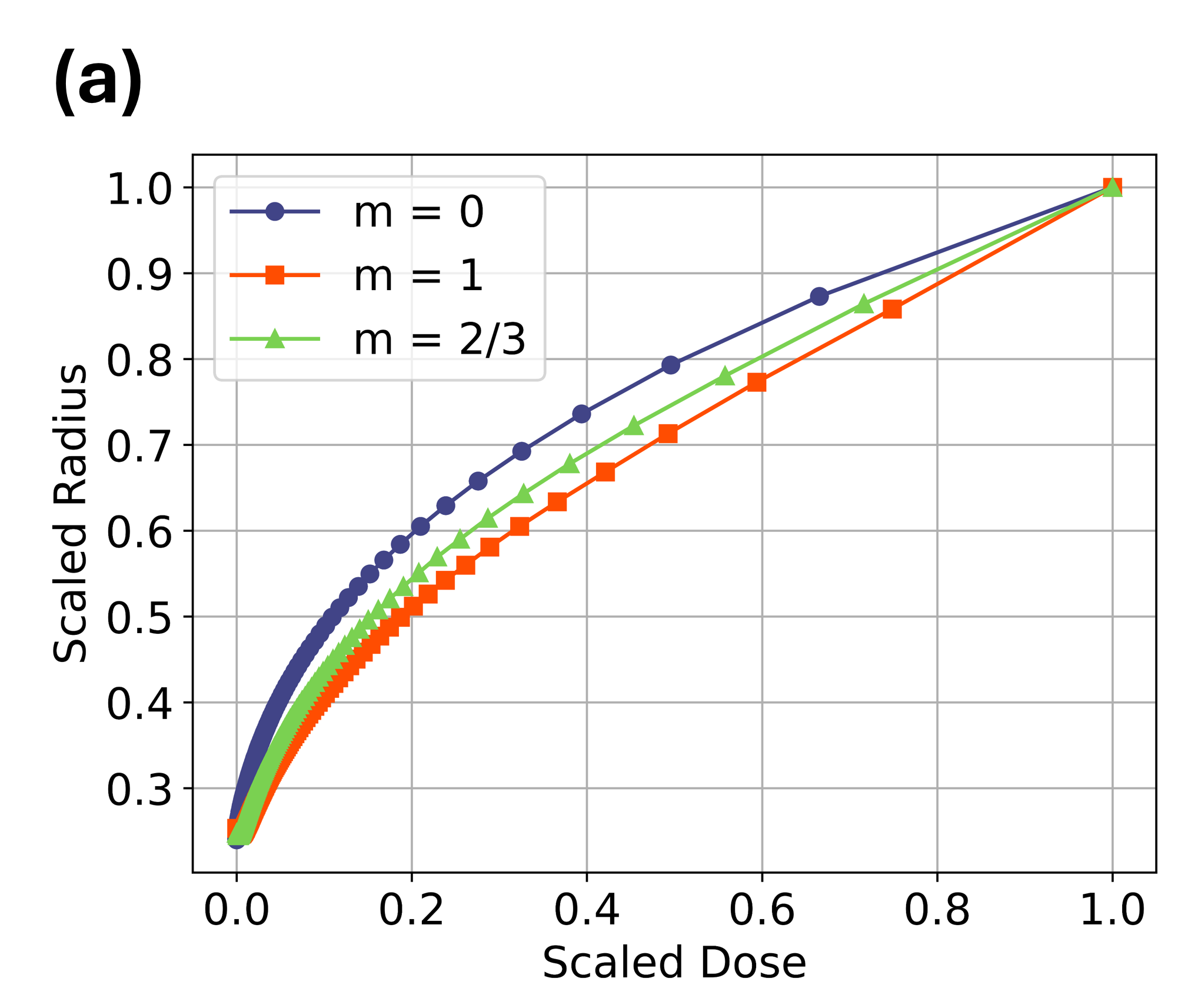}
\end{subfigure}
\begin{subfigure}{0.5\textwidth}
\includegraphics[width=\linewidth]{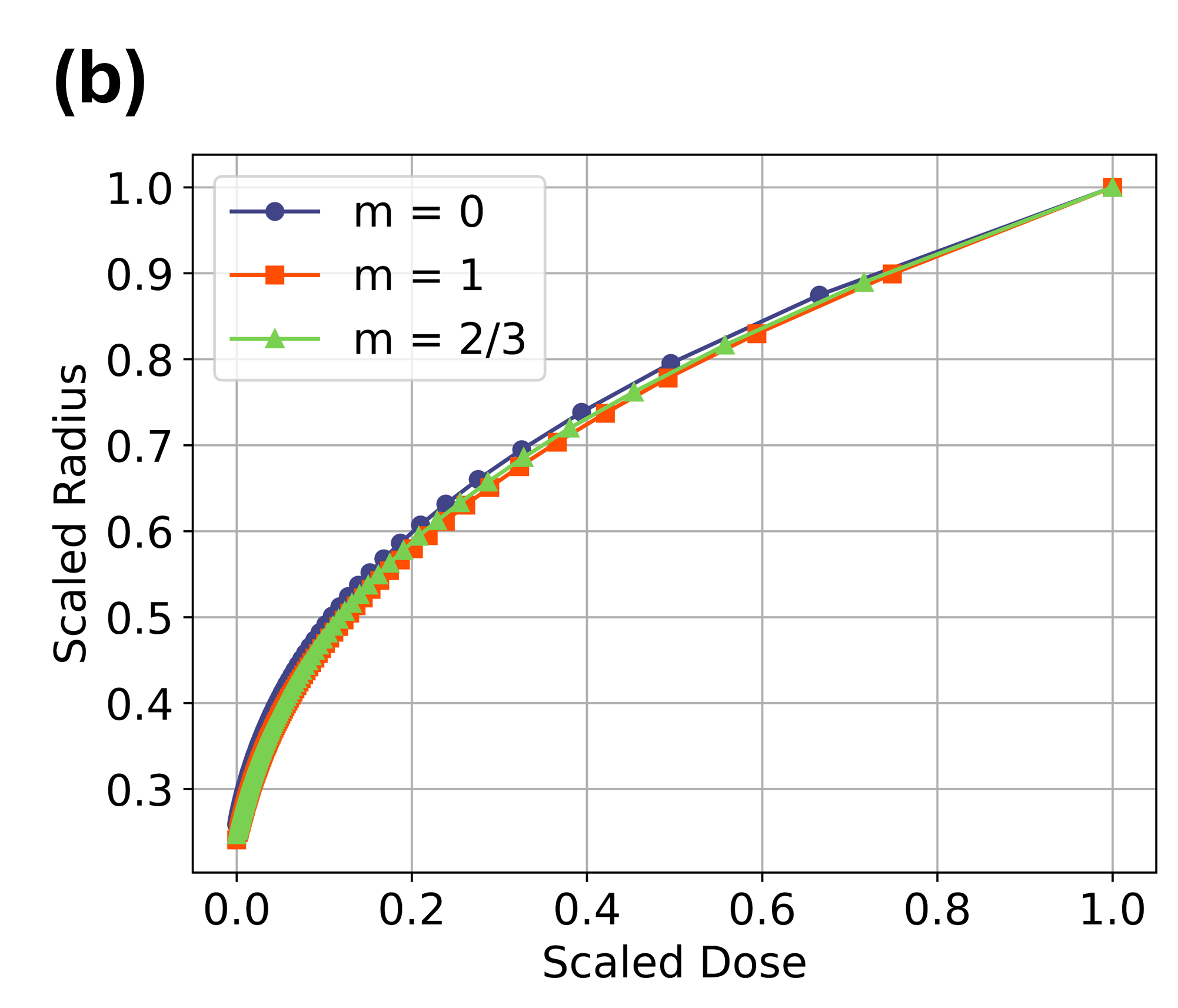}
\end{subfigure}
\caption{Scaled dose against scaled radius for the radius increase computational model; (a) Average radius, and (b) Volume weighted average radius.} \label{fig:Fig18}
\end{figure}

To determine why the radius appears to increase linearly with dose, a computational model was created to explore increase in grain radius with dose based on selection criteria fitting certain hypotheses. Since the grain growth is fitted to the sheared cells, these are used as a starting point for the model. We start with the same cell volume as the MD simulation cells and divide this by 135 grains, which is the average number of grains at low dose in the sheared cells, shown in Figure \ref{fig:Fig6}(a).

As with the Toy Model in the Discussion section, the grains are allowed to exchange volume until a critical minimal volume value is reached and a grain disappears. A gaussian distribution is sampled from to facilitate the exchange and the volume exchange is biased based on different hypotheses. From the grain volumes, the mean grain radius can be extracted. 

The hypotheses that bias the selection are as follows:
\begin{itemize}
    \item $m=0$ states that every grain is equally likely to grow/shrink.
    \item $m=1$ states that the growth rate is proportional to the volume of the grain.
    \item $m=2/3$ states that grain growth rate is proportional to grain surface area.
\end{itemize}

A plot of scaled radius against scaled dose is shown in Figure \ref{fig:Fig18}(a). This figure shows the scaled average grain radius as a function of scaled dose

\begin{equation} \label{eqC2}
    \langle r \rangle = \Sigma_{i}r_{i}/N
\end{equation}

where N is the number of grains. However, the WH analysis of the MD data generates a volume average grain radius given by

\begin{equation} \label{eqC3}
    \langle r \rangle_{vol} = \Sigma_{i}V_{i}r_{i}/\Sigma_{i}V_{i} = \langle r^{4} \rangle / \langle r^{3} \rangle
\end{equation}

This volume weighted average is shown in Figure \ref{fig:Fig18}(b). Whilst this figure is not perfectly linear, the linear term dominates, consistent with the MD results.

\section{Toy Model Convergence Study}

\begin{table}
  \caption{Toy Model convergence study data for each proposed exchange mechanism.}
  \label{tbl:Table2}
  \includegraphics[width=\linewidth]{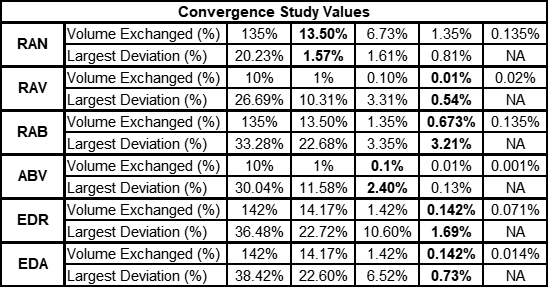}
\end{table}

It is important that the volume exchanged between grains is as low as possible between steps whilst also keeping the calculation time reasonable in the Toy Model. Hence, the Toy Model was run for each distinct exchange mechanism until an acceptable level of convergence was obtained. The d$_{c}$ = 5.2 nm cell case was chosen for the comparisons which relates to an average grain volume of $\sim$ 74 nm$^{3}$ which is used to determine the exchange percentage. 

Table \ref{tbl:Table2} shows the results of the convergence study for the distinct simulations. For each distinct exchange mechanism, there are five listed volume exchange percentages which show the percentage of volume exchanged between the two grains in each step. The deviation percentage below shows the largest percentage difference in values between the ranks of the current and next smaller exchange percentage e.g. the largest deviation between exchanging 10\% and 1\% volume for the RAV case is 26.69\%. This large difference indicates a lack of convergence. Where a rank disappears between smaller exchange volumes, an additional check is carried out whereby the overall value of grains disappearing at that rank cannot exceed the largest deviation between any other ranks i.e. if rank 10 disappears in between runs, so the 10th smallest grain no longer disappears next, then the overall percentage value of that rank cannot exceed the largest deviation in the subsequent run.

In Table \ref{tbl:Table2}, the volumes which were ultimately exchanged between steps are highlighted in bold. The largest deviation is found to be 3.21\% for the RAB mechanism. The other deviations are much lower with some being as low as 0.54\%. These values were accepted to represent a converged model when considering computational efficiency. 

 \bibliographystyle{elsarticle-num} 
 \bibliography{reference}

\end{document}